\documentclass[a4paper,fleqn,usenatbib,useAMS]{mnras}

\usepackage{graphicx}   
\usepackage{amsmath}    
\usepackage{amssymb}    
\usepackage{multicol}        
\usepackage{bm}         
\usepackage{pdflscape}  
\usepackage{comment}

\usepackage[T1]{fontenc}
\usepackage{ae,aecompl}

\title[Optical and IR variability stduy of I19520]{Testing the bloated star hypothesis in the massive young stellar object IRAS 19520+2759 through optical and infrared variability}
\author[Pandey et al.]
{Rakesh Pandey$^{1}$\thanks{pandey.rakesh405@gmail.com},  
Aina Palau$^{1}$,
Javier Serna$^{2,3}$,
Rolf Kuiper$^{4}$,
Alvaro S\'anchez-Monge$^{5,6}$
\newauthor
Saurabh Sharma$^{7}$,
Raghvendra Sahai$^{8}$,
Carmen S\'anchez Contreras$^{9}$,
Jes\'us Hern\'andez$^{2}$,
\newauthor
Carlos Rom\'an-Z\'u\~niga$^{2}$,
Florian Rodler$^{10}$
\\
$^{1}$Universidad Nacional Aut\'onoma de M\'exico, Instituto de Radioastronom\'ia y Astrof\'isica, Antigua Carretera a P\'atzcuaro 8701,\\ 
Ex-Hda. San Jos\'e de la Huerta, 58089 Morelia, Michoac\'an, M\'exico\\
$^{2}$Universidad Nacional Aut\'onoma de M\'exico, Instituto de Astronom\'ia, AP 106, Ensenada 22800, BC, M\'exico\\
$^{3}$Homer L. Dodge Department of Physics and Astronomy, University of Oklahoma, Norman, OK 73019, USA \\
$^{4}$Faculty of Physics, University of Duisburg-Essen, Lotharstra{\ss}e 1, D-47057 Duisburg, Germany \\
$^{5}$Institut de Ci\`encies de l'Espai (ICE, CSIC), Campus UAB, Carrer de Can Magrans s/n, 08193, Bellaterra (Barcelona), Spain \\
$^{6}$Institut d'Estudis Espacials de Catalunya (IEEC), 08860 Castelldefels (Barcelona), Spain \\
$^{7}$Aryabhatta Research Institute of Observational Sciences (ARIES), Manora Peak, Nainital, 263 002, India \\
$^{8}$Jet Propulsion Laboratory, MS 183-900, California Institute of Technology, Pasadena, CA 91109, USA\\
$^{9}$Centro de Astrobiolog'ia (CAB), CSIC-INTA, ESAC, Camino Bajo del Castillo s/n, E-28692, Villanueva de la Ca\~nada, Madrid, Spain \\
$^{10}$European Southern Observatory, Alonso de Cordova 3107, 7630391 Vitacura, Santiago de Chile, Chile \\
}

\date{Accepted XXX. Received YYY; in original form ZZZ}

\pubyear{2023}

\newcommand{\msunyr}{{M}_{\sun}~{\rm yr}^{-1}}
\newcommand{\mdot}{\dot{M}_{\ast}}
\newcommand{\msun}{{M}_{\sun}}
\newcommand{\rsun}{{R}_{\sun}}
\newcommand{\lsun}{{L}_{\sun}}

\newcommand{\js}[1]{{\color{green} #1}}

\begin{document}
\label{firstpage}
\pagerange{\pageref{firstpage}--\pageref{lastpage}}
\maketitle

\begin{abstract}
{
Using optical time series with Telescopi Joan Or\'o (TJO), Gaia, TESS, and NEOWISE archival data, we performed a variability study on the candidate bloated massive young stellar object (MYSO) IRAS 19520+2759. This is the first time that a bloated star candidate has been tested for the theoretically predicted periodic variability. The source is found to be variable at optical and mid-infrared wavelengths and classified as a long-period variable MYSO. The observed TJO data gives a period of the source of $\sim$ 270$\pm$40 days (in the Rc band) and $\sim$ 270$\pm$50 days (in the Ic band), which is very close to the value predicted by the theoretical Period-Luminosity relation for a bloated young star of $\sim10^5$~$\lsun$. 
Additionally, a large period of $\sim$ 460$\pm$80 days (in the G band) and $\sim$ 440$\pm$70 (in the Rp band) is also visible in the Gaia light curve. 
The physical parameters of the source, such as mass, radius, and accretion rate, based on the theoretical predictions for the spherical accretion case and corresponding to a period of 270--460 days, are $\sim 24$--28$\,\msun$, $\sim 650$--900$\,\rsun$ and $\sim (6$--$9)\times10^{-3}\,\msunyr$.
However, these numbers are very sensitive to the effective temperatures assumed in the models. Additionally, these values strongly depend on the geometry of accretion and could significantly decrease 
for the case of a MYSO accreting 
through a disc.
The observed periodic variability, the observed colour trend, and the nature of the variability are found to be consistent with the pulsational model for a bloated MYSO. }
\end{abstract}


\section{Introduction} \label{sec1}

Massive stars (M $>$ 8 $\msun$) are thought to be formed at the center of hub-filament systems,
accreting at a rate of $> 10^{-4} \, \msunyr$ \citep{krum2009,kuper16,Kumar2020}, with the accretion history of the star playing an important role in its evolution \citep{yb08,kuiper13}.
Although several theoretical and observational works have shed light on the formation process and the initial evolution of massive stars \citep{bonnell01,McKee2003,tige17,motte2018}, the details of these processes are far from completely understood. An important aspect, robustly predicted by theory, is that a massive young stellar object (MYSO) accreting at such accretion rates should bloat or swell up \citep{Hosokawa2009,kuiper13,hle2016}. The increased radius due to bloating results in a lower effective temperature and lower UV luminosity of the MYSO. \citet{lumsden2013} attributed the lack of HII regions around some MYSOs with high bolometric luminosity ($\sim$ 10 $^{5}$ $\lsun$) to the bloating due to high accretion rates. Thus, two observational properties are critical to consider an object as a bloated star candidate: centimetre emission fainter than expected and a spectral type later than expected given its measured luminosity, assuming that the object is at the Zero-Age Main Sequence (ZAMS) \citep{Hosokawa2009,hoskawa2010}. 
This consideration seems reasonable since the bloated phase is supposed to occur at the end of the protostellar phase when the object is about to approach the ZAMS, and the luminosity is expected to be of the order of the ZAMS luminosity (e.g., \citealt{hoskawa2010}).

Other observational diagnostics of the bloated phase of a massive star come from theoretical works like those of \citet{inayoshi2013}. They performed linear stability analysis on a rapidly accreting massive protostar and predicted that these sources become pulsationally unstable during the bloating phase. They also suggested that this pulsational instability causes periodic variability in these sources (depending on the accretion rate), and came up with a period-luminosity (PL) relation given by $\mathrm{log}(L/\lsun) = 4.62 + 0.98\,\mathrm{log}(P/100\,\mathrm{days})$. The authors further derived equations to determine the mass, radius, and accretion rate based on the period of the bloated source. 
This study provides a potential methodology to characterise this type of sources through their variability.

IRAS 19520+2759 (I19520; RA (J2000) =19h54m05.9s, Dec(J2000)=+28d7m41s) is a candidate bloated MYSO \citep{Palau2013}. The source lies close to the boundary of the supernova remnant G65.1+0.6, and is embedded in an extended massive millimetre clump. Considering the radial velocity of the source ($V_\mathrm{LSR} \sim -16.5$\,km\,s$^{-1}$), \citet{Palau2013} derive a kinematic distance of 8.4 kpc. The photogeometric distance of the optical counterpart based on the Gaia EDR3 is reported as $\sim$ 9.1$\pm$1.6 kpc, whereas the geometric distance is given as $\sim$ 10.8$\pm$2.8 kpc \citep{bj21}, which are both close to the kinematic distance. Here we adopt a distance of 9 kpc. 

I19520 was initially recognised as an OH/IR star \citep{Habing84}. Later, a far-infrared (FIR) study and spectral energy distribution (SED) analysis revealed the source to be a MYSO \citep{Hrivnak1985}. This source is one of the YSOs with highest bolometric luminosity ($\sim$ $10^{5}$ $\lsun$) in the Red MSX Source (RMS) survey (cf. Table 2, \citealt{lumsden2013}). The optical counterpart of this source, identified by \citet{Habing84}, is quite bright (mean G mag of $\sim$ 14.9) and was detected as a point-like source in Hubble Space Telescope images. Using optical spectra, \citet{conteras2008} identified the spectral type of the source as O9.5V or early B. 
The multiwavelength study of the source performed by \citet{Palau2013} identified it as an actively accreting MYSO driving a collimated outflow. The authors designated the source I19520 as a candidate bloated MYSO due to its tentative spectral type, later than expected for a 10$^5$~$\lsun$ ZAMS star, the absence of free-free emission, and the accretion properties of the millimetre source. 

For the case of I19520, the PL relation given by \citet{inayoshi2013} yields a period of 244 days. To test if I19520 follows the aforementioned PL relation, we carried out a variability study of the source at optical and infrared wavelengths and complemented it with archival data sets.
We have organized the manuscript as follows: 
Section \ref{sec2} gives details and procedures regarding the observations, image reduction, and archival data sets. In Section \ref{sec3}, we present the analysis of the different data sets and important outcomes. We discuss the results in Section \ref{sec4}, and the conclusions are summarised in Section \ref{sec5}.

\begin{figure*}
\centering
\includegraphics[width=.45\textwidth]{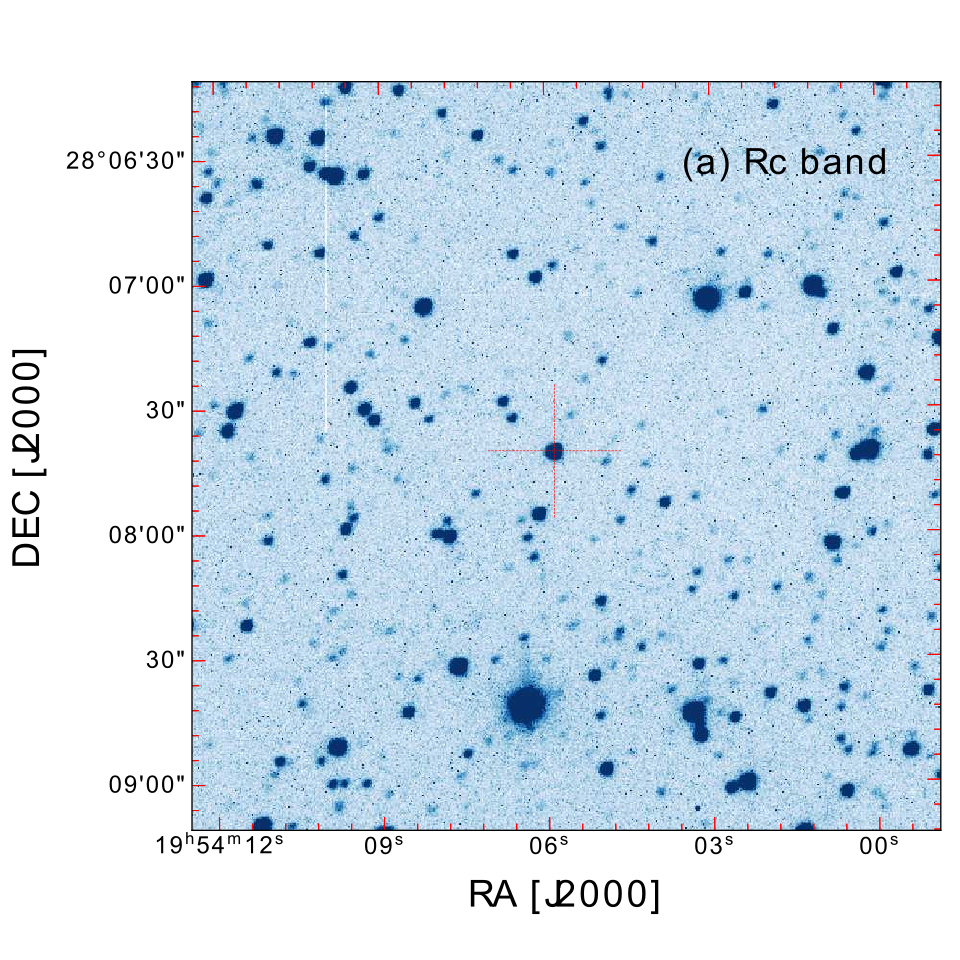}
\includegraphics[width=.45\textwidth]{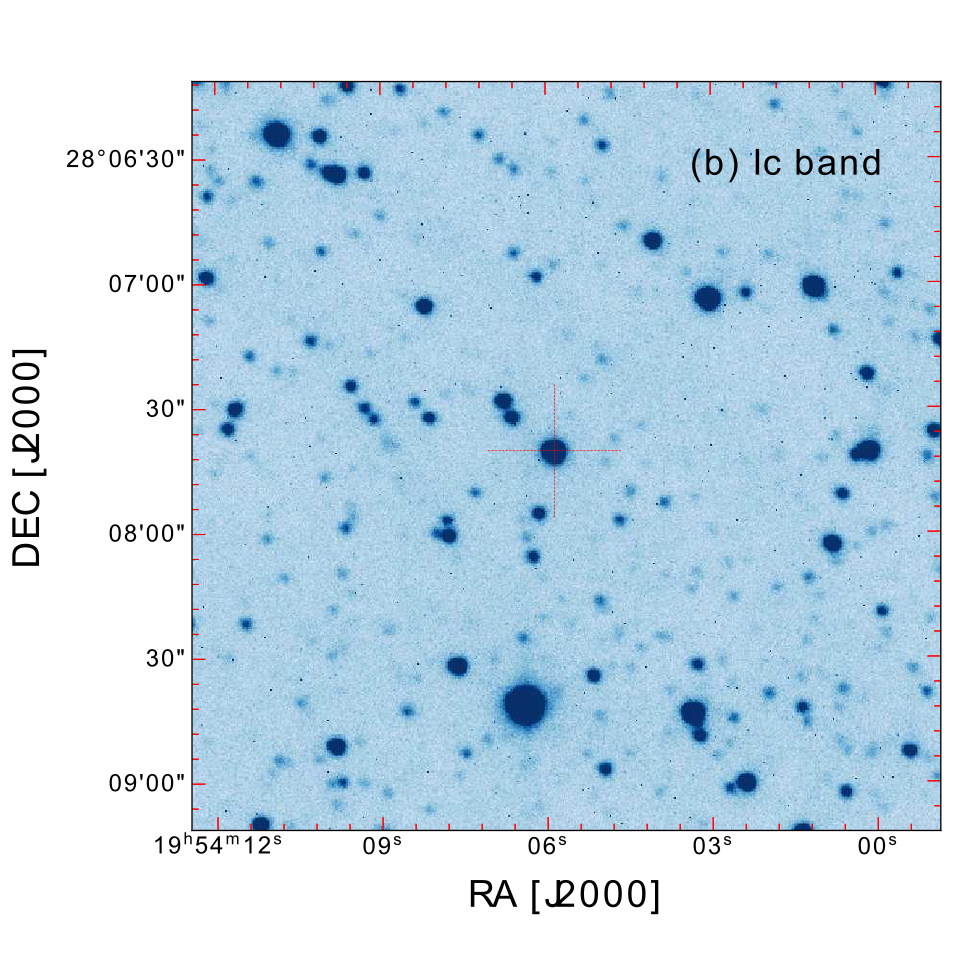}
\caption{\label{fig1} TJO Rc (a) and Ic (b) band images of the region surrounding I19520. The red cross on both panels marks the position of the target.}
\end{figure*}

\begin{table*}
\centering  
\caption{TJO observation log including observing date, number of frames multiplied by the exposure time, and filters used.}
 \label{tab1}
 \begin{tabular}{|p{5cm}|p{5cm}||p{3cm}|}
  \hline
Date of observation        & N$\times$Exp.(s)     &  Filters\\
  \hline
\\
2013.05.25             &   13$\times$120         & $I_{c}$     \\
2013.05.26             &   13$\times$120         & $I_{c}$     \\
2013.06.02             &    3$\times$300        & $R_{c}$     \\
2013.06.03             &    15$\times$300, 43$\times$120         & $R_{c}$, $I_{c}$     \\
2013.06.04             &    24$\times$300, 46$\times$120         & $R_{c}$, $I_{c}$     \\
2013.06.05             &    24$\times$300, 30$\times$120         & $R_{c}$, $I_{c}$     \\
2013.06.12             &    9$\times$300, 15$\times$120         & $R_{c}$, $I_{c}$     \\
2013.06.13            &    9$\times$300, 30$\times$120         & $R_{c}$, $I_{c}$     \\
2013.07.29            &    12$\times$300, 30$\times$120         & $R_{c}$, $I_{c}$     \\
2013.07.30            &    12$\times$300, 30$\times$120         & $R_{c}$, $I_{c}$     \\
2013.07.31            &    6$\times$300, 15$\times$120         & $R_{c}$, $I_{c}$     \\
2013.08.05            &    18$\times$300, 37$\times$120         & $R_{c}$, $I_{c}$     \\
2013.08.06            &    18$\times$300, 53$\times$120         & $R_{c}$, $I_{c}$     \\
2013.12.01            &    12$\times$300, 30$\times$120         & $R_{c}$, $I_{c}$     \\
2014.04.29            &   12$\times$300, 30$\times$120          & $R_{c}$,$I_{c}$  \\
2014.05.02              &   12$\times$300,  30$\times$120         & $R_{c}$,$I_{c}$ \\
2014.05.03              &   12$\times$300,  30$\times$120         & $R_{c}$,$I_{c}$   \\
2014.05.04              &   12$\times$300,  30$\times$120         &  $R_{c}$,$I_{c}$  \\
2014.05.05              &   12$\times$300, 30$\times$120      & $R_{c}$,$I_{c}$  \\
2014.05.27              &   12$\times$300, 30$\times$120        & $R_{c}$,$I_{c}$   \\
2014.05.31              &   01$\times$300                     &  $R_{c}$   \\
2014.06.01              &   10$\times$300, 30$\times$120     & $R_{c}$,$I_{c}$   \\
2014.06.20              &   12$\times$300, 20$\times$120     & $R_{c}$, $I_{c}$    \\
2014.06.21              &   07$\times$120                      &    $I_{c}$\\
2014.06.26              &   09$\times$300, 27$\times$120       &    $R_{c}$, $I_{c}$    \\
2014.06.27              &   12$\times$300, 30$\times$120       &     $R_{c}$,$I_{c}$   \\
2014.07.05              &   12$\times$300, 30$\times$120     &  $R_{c}$,$I_{c}$  \\
2014.07.08              &   11$\times$300, 15$\times$120     & $R_{c}$,$I_{c}$   \\
2014.07.09              &   1$\times$300, 14$\times$120      & $R_{c}$,$I_{c}$   \\
2014.07.11              &   11$\times$300, 15$\times$120      & $R_{c}$,$I_{c}$   \\    
2014.07.12              &   01$\times$300, 15$\times$120     & $R_{c}$,$I_{c}$   \\     
2014.07.20              &   12$\times$300, 30$\times$120    & $R_{c}$,$I_{c}$   \\ 
2014.07.29              &   11$\times$300, 15$\times$120      & $R_{c}$,$I_{c}$   \\
2014.07.30              &   01$\times$300, 15$\times$120    &  $R_{c}$,$I_{c}$  \\
2014.07.31              &   11$\times$300, 15$\times$120    &  $R_{c}$,$I_{c}$   \\
2014.08.01              &   01$\times$300, 15$\times$120     & $R_{c}$,$I_{c}$   \\
2014.08.24              &   12$\times$300, 30$\times$120   & $R_{c}$,$I_{c}$  \\
2014.08.25              &   12$\times$300, 30$\times$120     & $R_{c}$,$I_{c}$  \\
2014.08.26              &   12$\times$300, 30$\times$120     & $R_{c}$,$I_{c}$  \\
2014.08.28              &   12$\times$300, 27$\times$120     & $R_{c}$,$I_{c}$  \\
2014.09.01              &   12$\times$300, 30$\times$120    & $R_{c}$,$I_{c}$ \\
2014.09.02              &   12$\times$300, 30$\times$120    & $R_{c}$,$I_{c}$  \\
2014.09.10              &   12$\times$300, 30$\times$120    & $R_{c}$,$I_{c}$  \\   \hline

\end{tabular}
\end{table*}
\begin{figure*}
\centering
\includegraphics[width=0.45\textwidth]{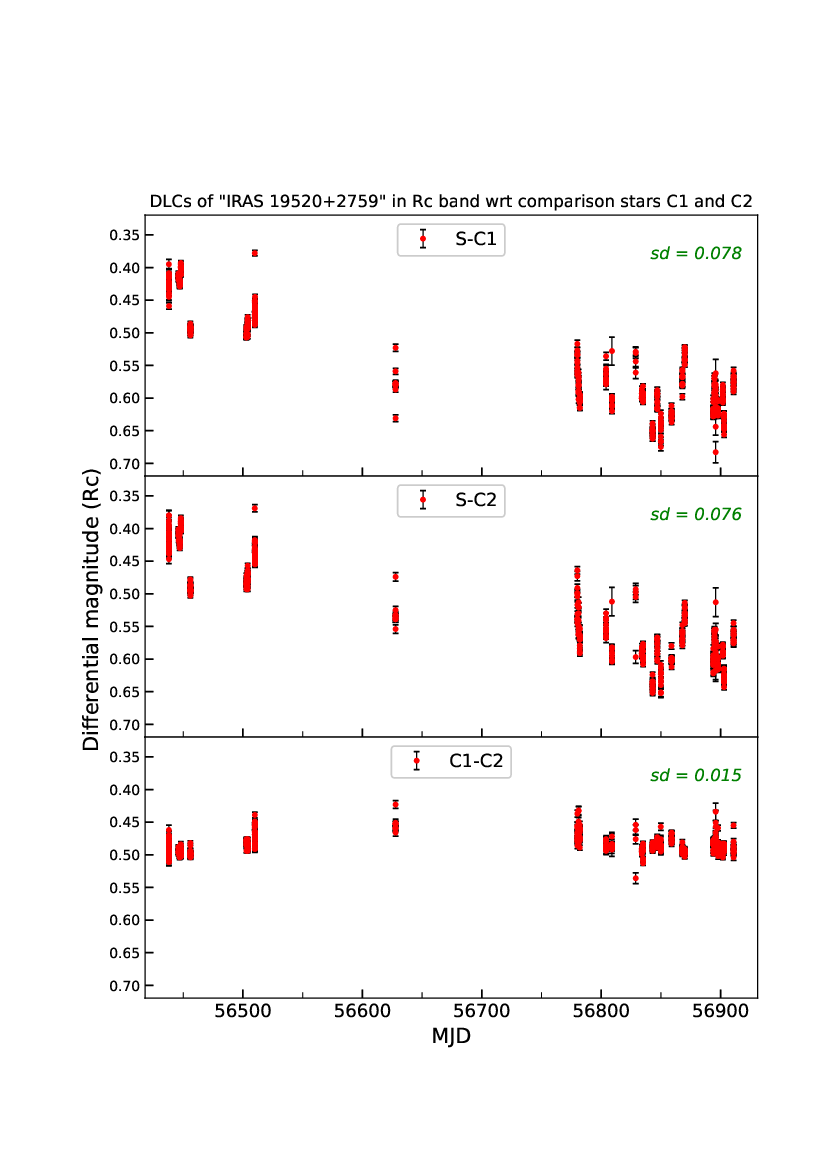}
\includegraphics[width=0.45\textwidth]{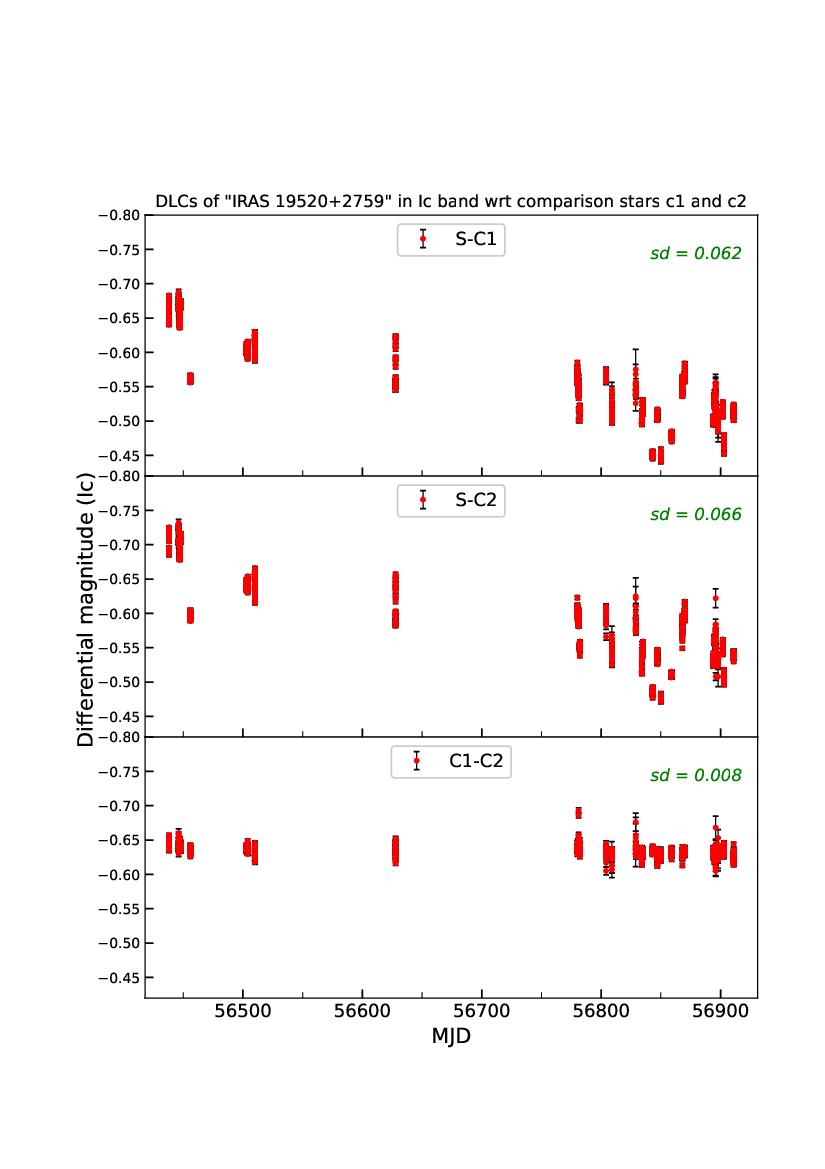}
\caption{\label{fig2} DLCs of I19520 in the TJO Rc (left panel) and Ic bands (right panel). The top and middle panels of both figures show the DLCs with respect to the comparison stars c1 and c2, and the bottom panel shows the difference between the comparison stars.}
\end{figure*}

\begin{figure*}
\centering
\includegraphics[width=0.48\textwidth]{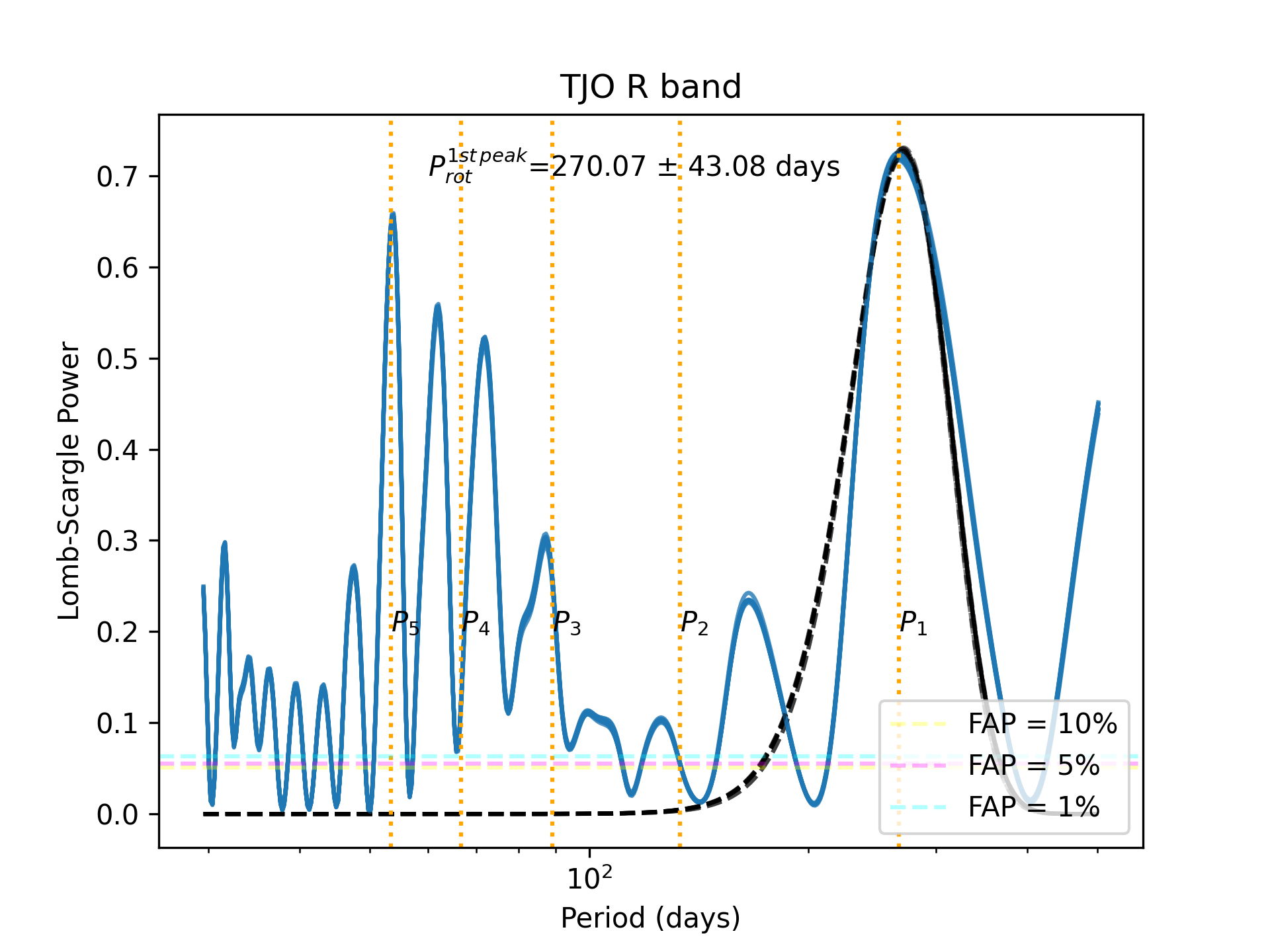}
\includegraphics[width=0.48\textwidth]{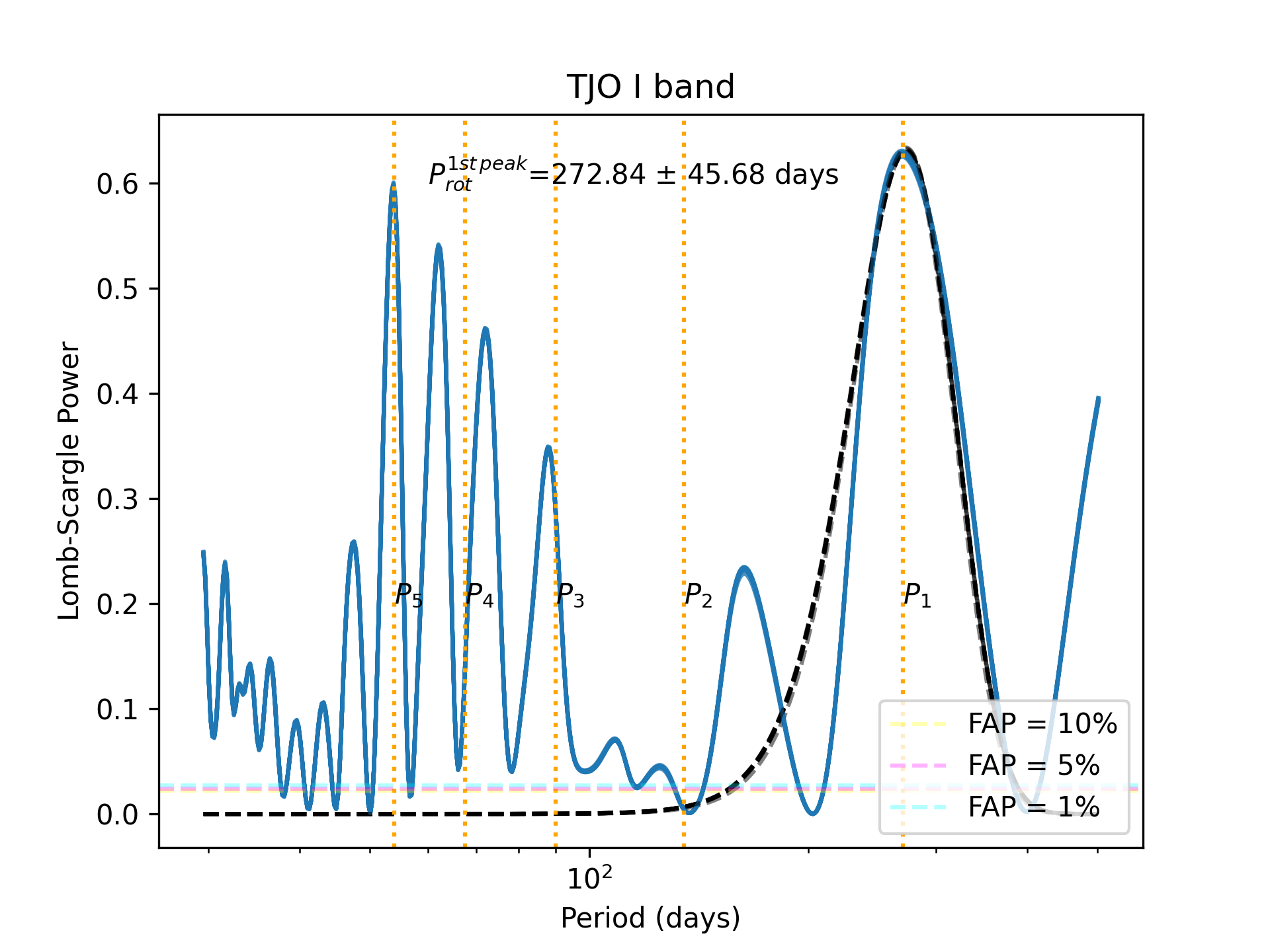}
\caption{\label{fig3} Lomb-Scargle Periodogram for the TJO Rc (left panel) and Ic (right panel) bands with False Alarm Levels of 1\%, 5\%, and 10\%, respectively. The orange dashed lines indicate the first five harmonics of the best period, which are estimated from the highest peak in the periodogram. The black dashed line is the Gaussian fit for the highest peak.}
\end{figure*}

\begin{figure*}
\centering
\includegraphics[width=1\textwidth]{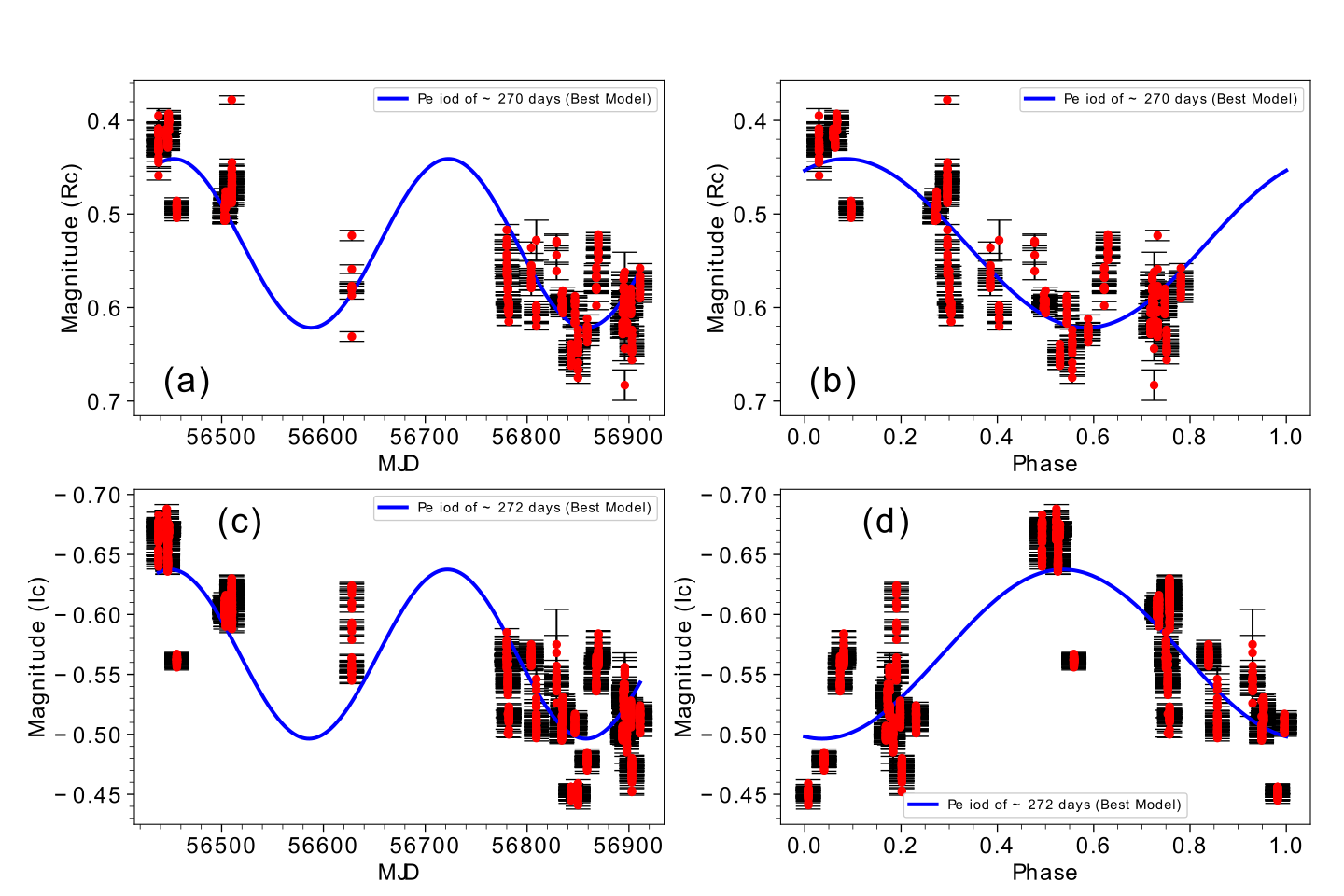}
\caption{\label{fig4} (a) DLC of I19520 in the TJO Rc band, overplotted with the best period of 270 days. (b) Phase-folded Rc band light curve of the source corresponding to the best period of 270 days. (c) DLC of the source in the TJO Ic band, overplotted with the best period of 272 days. (d) Phase-folded Ic band light curve of the source corresponding to the best period of 272 days.}
\end{figure*}

\section{Observations and archival data} \label{sec2}

\subsection{TJO data} \label{sec21}
The optical photometric data of the source in bands Rc and Ic were acquired using the `Telescopi Joan Or\'o' (TJO)\footnote{\url{https://montsec.ieec.cat/en/joan-oro-telescope/}} in Catalonia (Spain). The telescope, having a primary mirror diameter of 0.8~m, operates in a robotic mode. It is equipped with an optical imager named LAIA (Large Area Imager for Astronomy), which has a CCD of 4096 x 4096 pixels, and each pixel has a size of 0.4 arcseconds. This setup provides the telescope with a field of view (FOV) of 30$\times$30 arcmin in imaging mode. The source was observed in the optical Rc (6407\textup{~\AA}) and Ic (7980\textup{~\AA}) bands from May 25, 2013 to September 10, 2014, with multiple exposures taken at each epoch. The observation log of TJO is given in Table \ref{tab1}, and we show the Rc and Ic band images (single exposure) of the target in Figure \ref{fig1}.
We followed the usual procedure for cleaning the raw images as defined in \citet{rakesh2020,rakesh2022,pandey2023,rawat23}, and the frames taken during bad weather conditions were removed from the analysis. 
The frames in each filter were aligned using the Python package \texttt{Astroalign} \citep{astroalign19}. We performed aperture photometry on the cleaned images to get the magnitude of the source in each frame, using the \texttt{DAOPHOT} \citep{stetson87} package. We adopted the aperture radius as three times the full width half maximum (FWHM) of the source; the inner and outer sky annulus sizes were taken as 4$\times$FWHM+5 and 4$\times$FWHM+15, respectively. \\

\subsection{Archival data} \label{sec22}

To complement our observational data, we also used the publicly available optical epoch photometry data\footnote{\url{https://vizier.cds.unistra.fr/viz-bin/VizieR-3}} of our source from Gaia Early Data Release 3 \citep[DR3,][]{g21}. The output table contains transit-averaged Gaia magnitudes and fluxes (G, Bp, and Rp) along with the other parameters during a span of $\sim$ 1000 days. This work has additionally made use of data from the Transiting Exoplanet Survey Satellite \citep[TESS,][]{tess15}.
We also obtained multi-epoch MIR photometric data of the source in W1 (3.6 $\mu$m)
and W2 (4.5 $\mu$m) band from the Near-Earth Object Wide-field Infrared Survey Explorer (NEOWISE)\footnote{\url{https://irsa.ipac.caltech.edu/Missions/wise.html}}. The survey details and specifications are described in \citet{mainzer14}.

\section{Analysis} \label{sec3}

\subsection{TJO differential light curves and period analysis}  \label{sec31}
We generated differential light curves (DLCs) of the source in both filters to search for variations in the brightness of the source during the observed span and test for any potential variability.
The DLCs are proven to be a useful tool for inspecting the intrinsic variation in the brightness of astronomical sources \citep{tirth20, Mishra2021,joshi22,negi23}. This method eliminates spurious variations in the source caused by fluctuating sky conditions, the airmass effect, and instrumental artefacts.
To get the DLCs of the source, we need to get the differential magnitude of the source with respect to the comparison star (Target$-$Comparison) for all the frames, and plot it against time. For that, we made all the possible pairs among the stars in the magnitude bin ranging from (source-1) mag to (source+1) mag. We determined the difference in their magnitude for all the frames; the pair with the minimum standard deviation in their difference of magnitude is chosen as the best pair. As our final comparison star, we can select one of the sources from the best pair. We applied this method separately to both bands and determined the comparison stars accordingly. 

Finally, we plotted the differential magnitude against the MJD (Modified Julian Date) to get the DLCs of our source, see Figure \ref{fig2}.
The figure also shows the DLCs for the comparison stars (the best pair, as explained above). The variation in the differential magnitude of the target is quite large (with a standard deviation of $\sim$ 0.07 and $\sim$ 0.06 for the Rc and Ic bands, respectively) if we compare it with the scatter between the comparison stars (with a standard deviation of 0.015 and 0.008 for the Rc and Ic bands, respectively). The DLC of the source is similar among the best pair of comparison stars, indicating that the variation is caused by the target star. The DLCs of the source seem to present systematic variations (in both bands) in brightness when examined visually. The variation in the Rc band is $\Delta$Rc $\sim$ 0.3 mag, while in the Ic band is $\Delta$Ic $\sim$ 0.25 mag.

Since we visually identified systematic variations in the brightness of I19520, we tried to determine the period of the variation. We used the \texttt{Astropy} module \texttt{LombScargle} for the period estimation. The module is based on the Lomb–Scargle Periodogram (LSP) by \citet{Lomb1976} and \citet{Scargle1982}. 
Even when the data are not uniformly sampled and there are gaps in the data points, the LSP is still a valuable tool for estimating the period. We also estimated the uncertainty of the highest peak in the LSP following the Gaussian method described by \citet{Venuti_2017}. The uncertainty is quantified by the sigma value obtained from the Gaussian fitting process, which is performed using 20 points around the best period. The power spectra of I19520 in both TJO Rc and Ic bands, along with the uncertainty in the highest peak, are shown in Figure \ref{fig3}. The period with the highest power (the best period) is $\sim$ 270$\pm$40 days for the Rc band, and $\sim$ 270$\pm$50 days for the Ic band. 

In addition to the period uncertainty analysis, we plot the False Alarm Probability (FAP) levels for 1\%, 5\%, and 10\% as a reference to assess whether peaks are significant above the noise level. Peaks exceeding the 1\% FAP level are considered highly reliable, indicating that the signal has a 99\% confidence level, as per \citet{Scargle1982}. The periods of I19520 obtained for each TJO band are consistent with one another and are significantly higher than the 1\% FAP level (cf. Figure \ref{fig3}), hence they are highly reliable. We also plot the first five harmonics of the best period (P/2, P/3, P/4, and P/5) to identify and analyse if any dominant peak at the periodogram is an alias. Interestingly, the second best period (represented by the second highest peak in the periodogram) is found to be adjacent to the fifth harmonic (P/5) of the best period in both bands (cf. Figure \ref{fig3}), hence could be an alias. 

Figure \ref{fig4} (a, c) presents the DLC of I19520 in the TJO Rc and Ic bands, overplotted with the corresponding best period. The best periods fit reasonably well with the variation in the magnitude of the source during the whole observation period of $\sim$ 450 days. We then phase-folded the DLCs of the source with the corresponding best-periods, shown along with the best-period in Figure \ref{fig4} (b, d). The phase-folded light curves show a sinusoidal variation in the brightness of the source. Additionally, we cross-matched our estimated period values with the NASA Exoplanet Archive Periodogram service\footnote {\url{https://exoplanetarchive.ipac.caltech.edu/cgi-bin/Pgram/nph-pgram}}. The two approaches are in excellent agreement with one another.

\begin{figure*}
\centering
\includegraphics[width=0.48\textwidth]{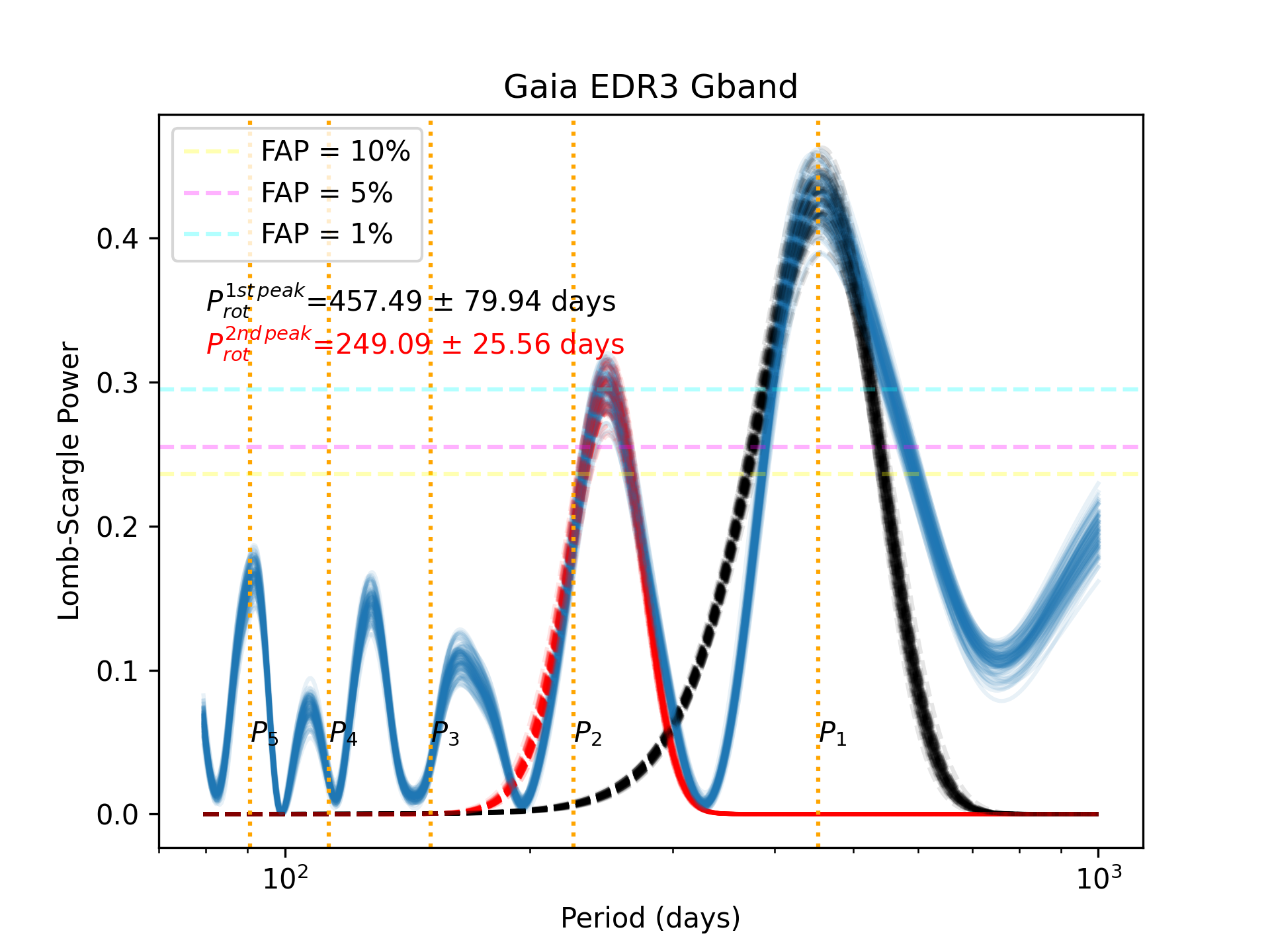}
\includegraphics[width=0.48\textwidth]{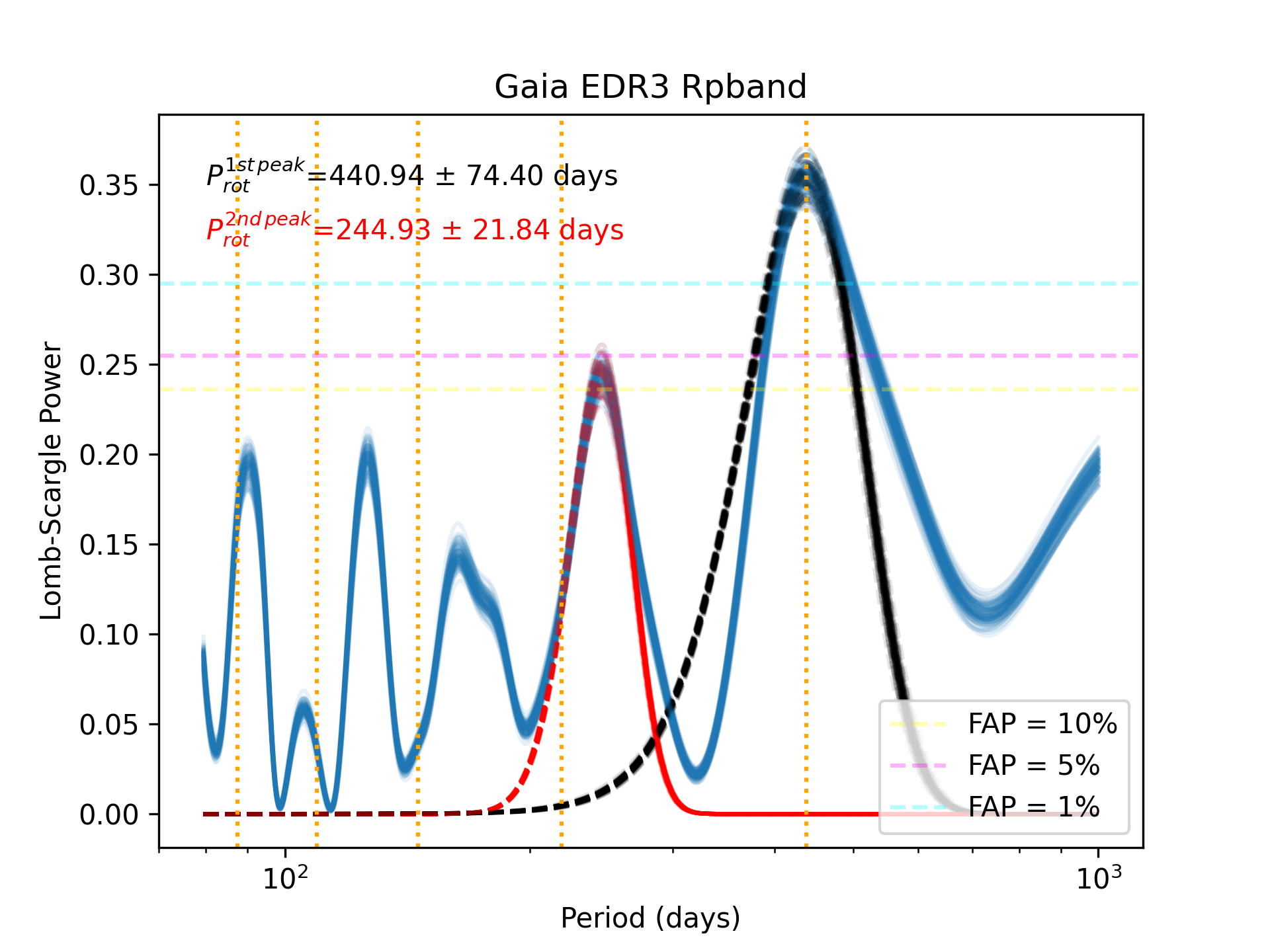}
\caption{\label{fig5} Lomb-Scargle Periodogram for Gaia G band (left panel) and Rp band (right panel) with False Alarm Levels of 1\%, 5\%, and 10\%, respectively. The orange dashed lines indicate the first five harmonics of the best period, which are estimated from the highest peak in the periodogram. Two Gaussian fits were applied to the two highest peaks, and the uncertainties are calculated based on the sigma values from the Gaussian fittings. }
\end{figure*}

\begin{figure*}
\centering
\includegraphics[width=.9\textwidth]{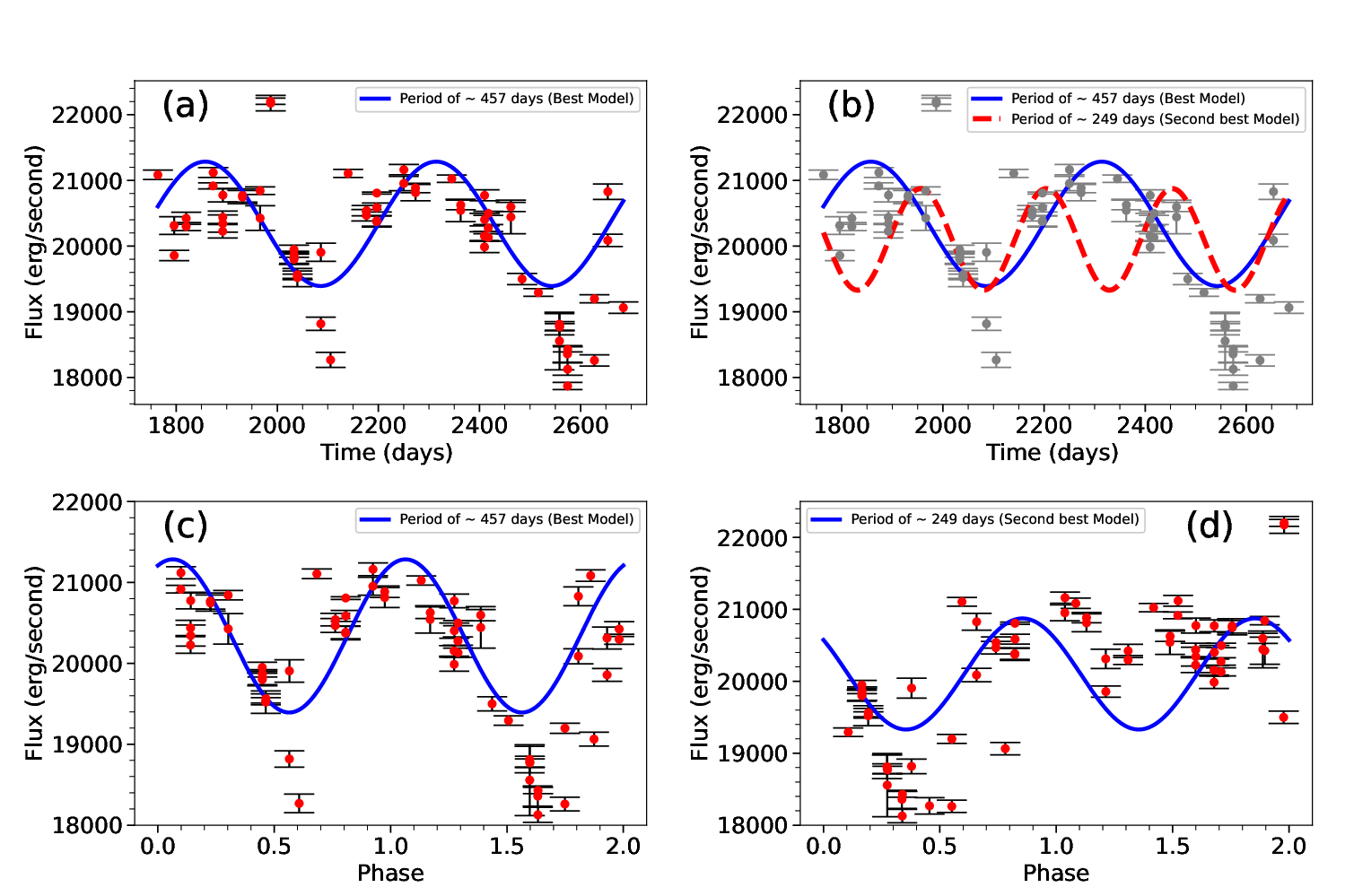}
\caption{\label{fig6} (a) Light curve of I19520 in the Gaia G band, overplotted with the best period of 457 days. (b) The same as previous panel but including also the second best period of 249 days. (c) Light curve phase-folded with a period of 457 days. (d) Light curve phase-folded with a period of 249 days.}
\end{figure*}

\begin{figure*}
\centering
\includegraphics[width=.9\textwidth]{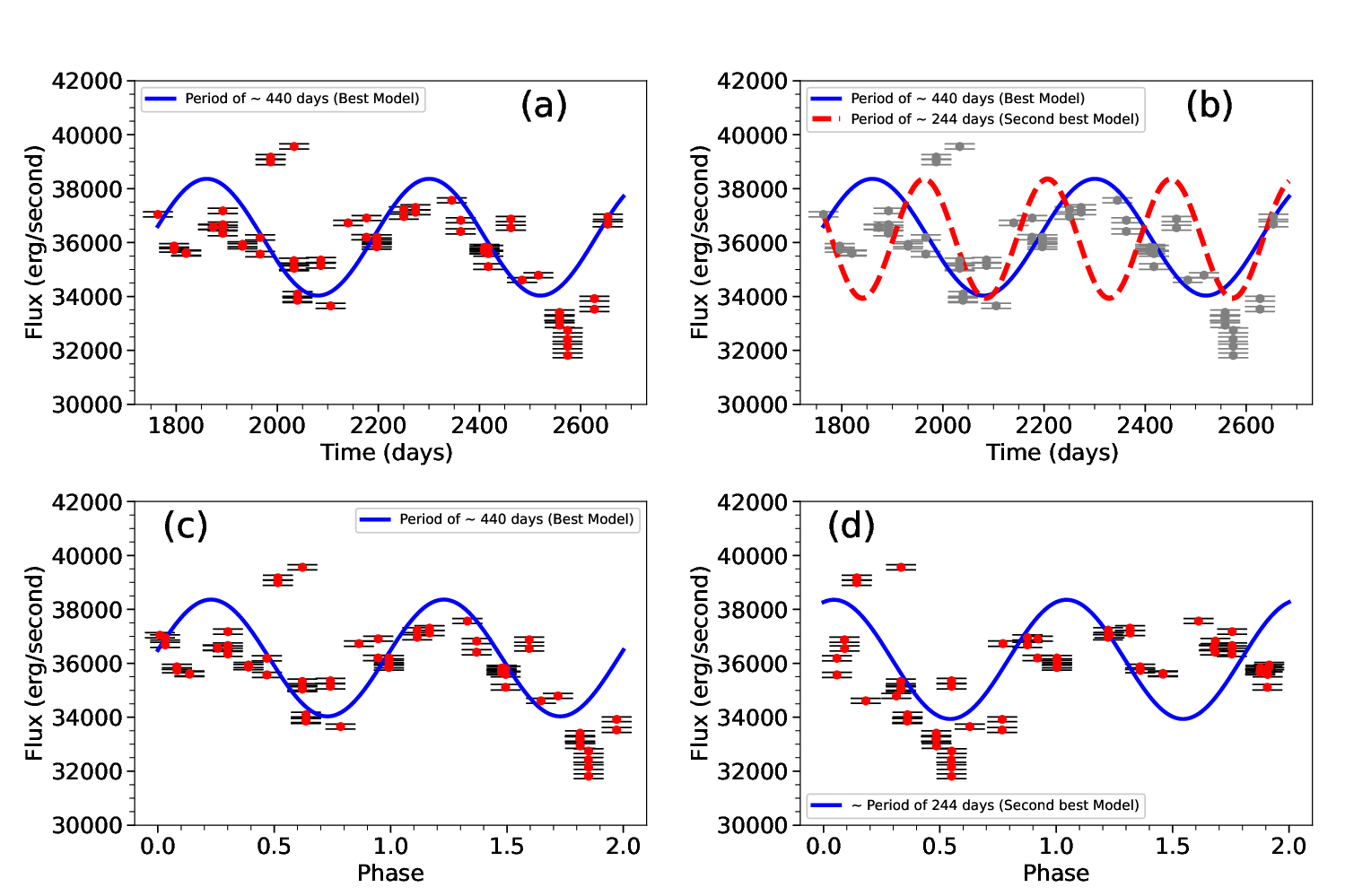}
\caption{\label{fig7} (a) Light curve of I19520 in the Gaia Rp band, overplotted with the best period of 440 days. (b) The same as previous panel but including also the second best period of 244 days. (c) Light curve phase-folded with a period of 440 days. (d) Light curve phase-folded with a period of 244 days.}
\end{figure*}

\begin{figure}
\centering
\includegraphics[width=.45\textwidth]{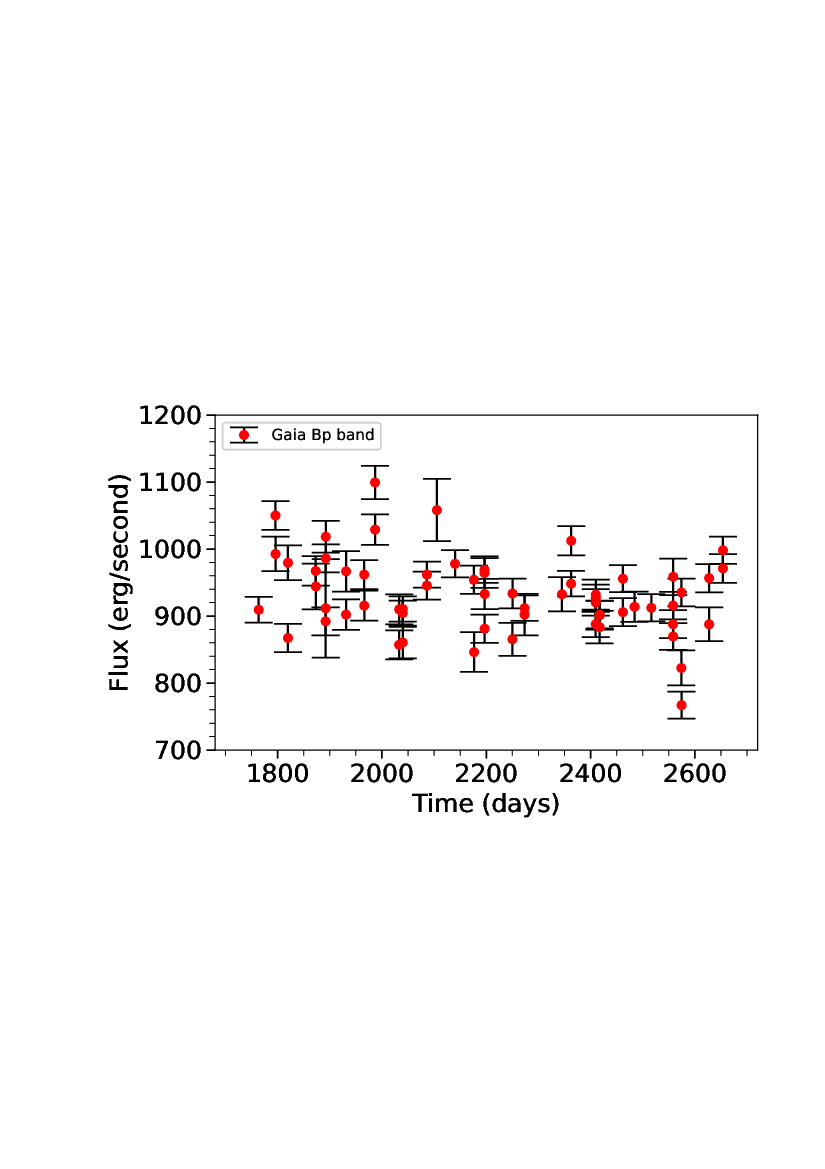}
\caption{\label{bp} Light curve of I19520 in the Gaia Bp band.}
\end{figure}

\subsection{Gaia light curves and period analysis} \label{sec32}

Additionally, we attempted to ascertain the target's period by performing a period and error analysis on the Gaia (G, Bp, and Rp) light curves (Figure \ref{fig5}). For the G band, the best and second best (corresponding to the second highest peak in the periodogram) periods are $\sim$ 460$\pm$60 and $\sim$ 250$\pm$30 days, respectively. Similarly for the Rp band, we have recovered the best period and second best period as $\sim$ 440$\pm$70 days and $\sim$ 240$\pm$20 days, respectively. However, we could not recover the same period for the Gaia Bp band data because the fractional errors in the Bp magnitude are significantly larger than for G and Rp bands (cf. Figure \ref{bp}).

The periods obtained using the Gaia G and Rp bands are consistent with each other and are above the 1\% FAP level (cf. Figure \ref{fig5}), thus seem quite reliable. For both bands, the second best period, however, is very close to the second harmonic (P2) of the best period (cf. Figure \ref{fig5}), considering the error in the periods.
Interestingly, a similar period of $\sim$ 270 days is also obtained using the TJO data (cf. Section \ref{sec31}). The TJO data have a 450-day total observation time, which is insufficiently long to enable a search for a $\sim$450 day period.

Figure \ref{fig6}-a presents the light curve of the target in the Gaia G band overplotted with the best period, and Fig. \ref{fig6}-b presents the same light curve overplotted with both the best and second best periods for a comparison. The phase-folded light curves with respect to the best period and second-best period, along with the corresponding periods, are also shown in the bottom panels of this figure. A similar plot for the Gaia Rp band is shown in Figure \ref{fig7}.
For both bands, the light curve phase-folded with the corresponding best period (Fig. \ref{fig6}-c and Fig. \ref{fig7}-c) shows a sinusoidal variation in the brightness of the source. However, it is worth noting that the Gaia G band light curve seems to deviate considerably from a smooth sinusoidal profile, especially its minima, which are deeper than the fitted sinusoidal curve (cf. Fig. \ref{fig6}-a). One explanation is that these might be noise or artefacts, although this does not seem likely given that both cycles appear to have deeper minima. Physical processes such as spot modulation, extinction variation, or small variation in mass accretion rates could potentially be causing this. These processes are usually responsible for scatters in the light curves of typical long-period YSOs, imposing small-scale scatters in a long-term periodic variation \citep{conteras16}. Given that the deeper minima are also visible at other wavelengths, the specific origin of these should be further examined. The phase-folded light curves corresponding to the second-best period significantly depart from a sinusoidal pattern (Fig. \ref{fig6}-d, Fig. \ref{fig7}-d). Thus, the best period in both bands seems more reliable than the corresponding second best period.

\begin{figure*}
\centering
\includegraphics[width=.8\textwidth]{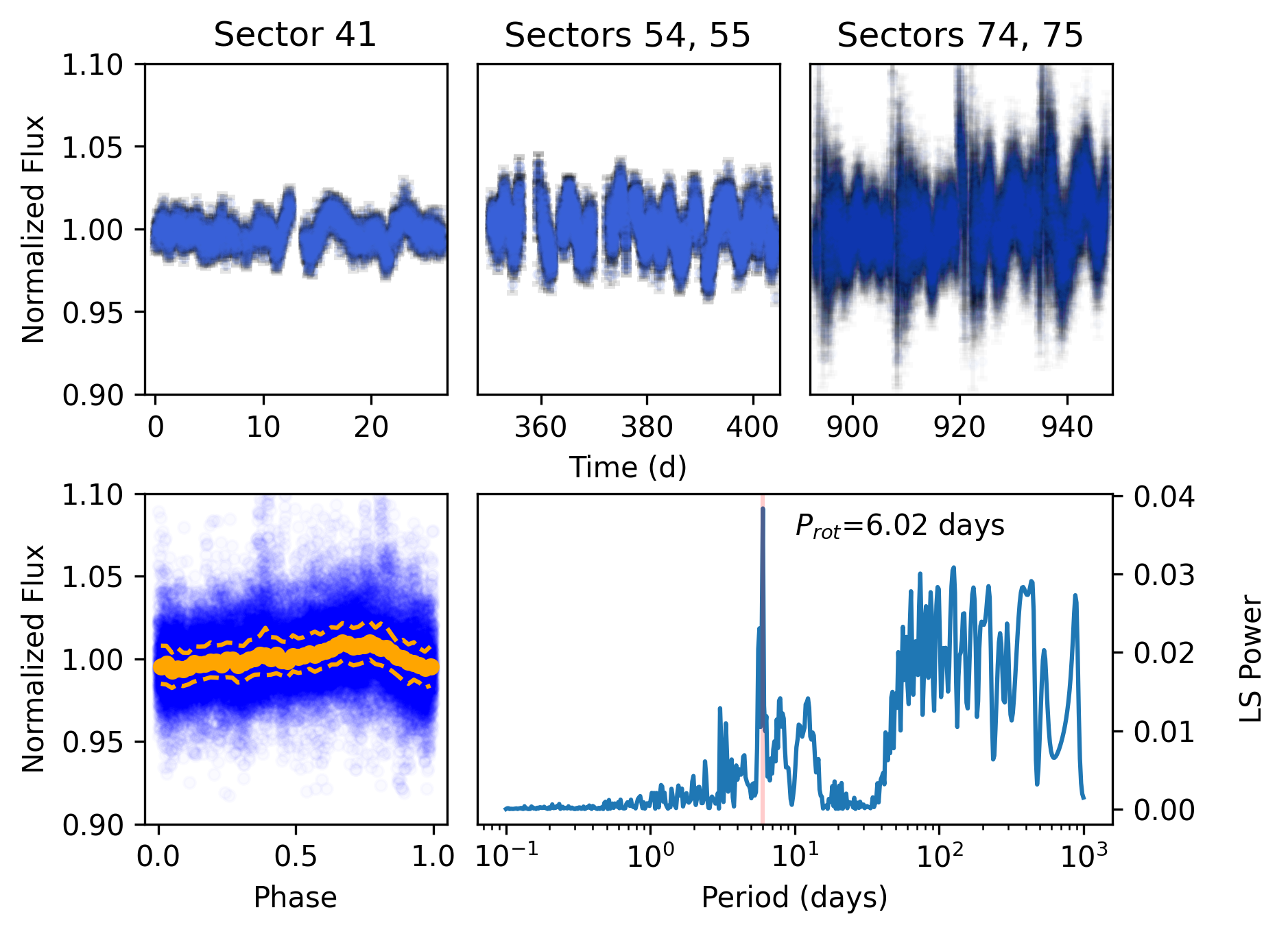}
\caption{\label{fig8} TESS light curves and period analysis. The top panels show the corrected light curves for all the TESS sectors. The bottom left panel shows the complete phase-folded light curve for the period of 6.02 days and the moving median and its interquartile range of the phase-folded light curve. The bottom right panel shows the Lomb-Scargle periodogram with the estimated period.}
\end{figure*}

\begin{figure*}
\centering
\includegraphics[width=.9\textwidth]{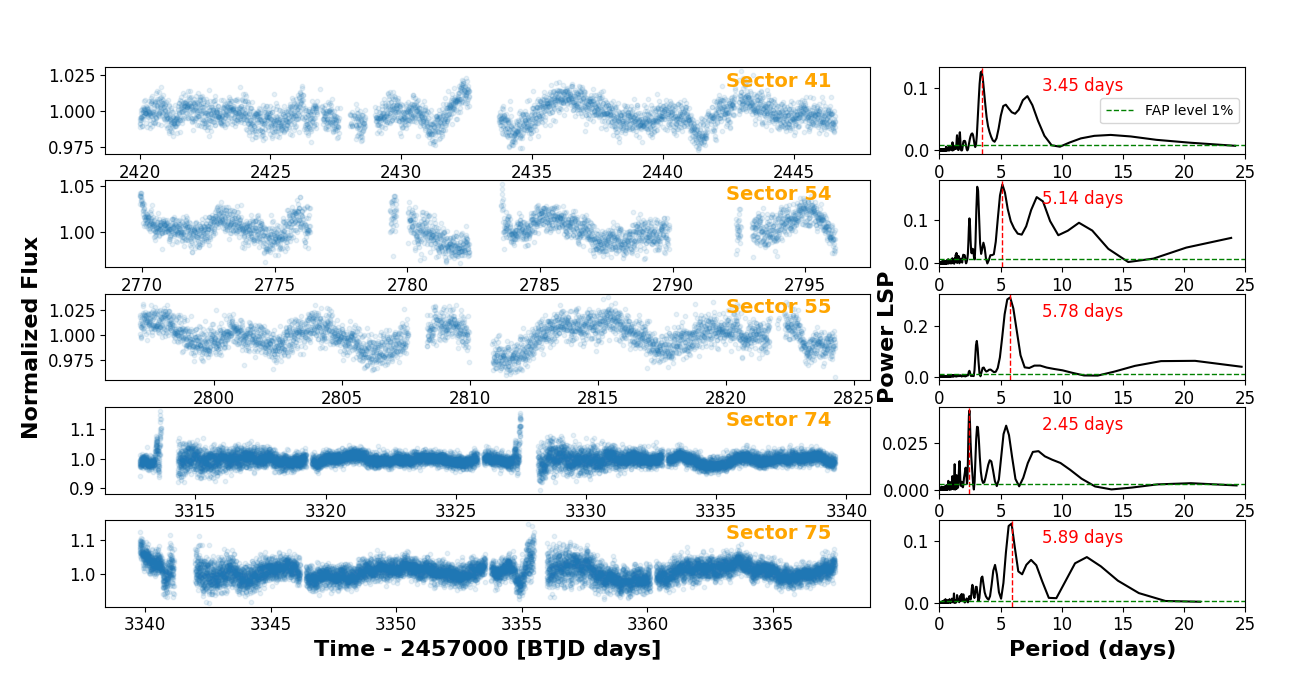}
\caption{\label{fig8n} TESS light curves corrected by systematic effects sector by sector (left panel) and Lomb-Scargle periodogram analysis (right panel). The red vertical line shows the estimated period at the highest peak in the periodogram, and the green dashed line displays the FAP at 1\% level. The best periods for three out of 5 sectors are very similar, around 5.6 days.}
\end{figure*}

\subsection{TESS light curves and period analysis} \label{sec33}

We used the web application \href{https://www.tessextractor.app}{\texttt{TESSExtractor}\footnote{\url{https://www.tessextractor.app}}} \citep{Serna_2021} to download data from the TESS and process light curves from full-frame images provided by the TESScut service \citep{TESSCut}. The application employs simple aperture photometry to extract stellar flux. Additionally, we used the single-scale Cotrending Basis Vectors (CBVs; \citealt{smith12}) from TESS and the \texttt{kepcotrend} task of the PyKE package \citep{PyKE} to correct the light curves for systematic effects. Each sector is corrected separately, optimizing the goodness of the fit between linear combinations of CBVs and the uncorrected light curve through the chi-squared value.
TESS has observed I19520 across multiple sectors, specifically: Sector 14 (July 18, 2019–August 15, 2019) [target is at the border of the CCD, data of bad quality], Sector 41 (July 23, 2021–August 20, 2021), Sector 54 (July 9, 2022–August 5, 2022), Sector 55 (August 5, 2022–September 1, 2022), Sector 74 (January 3, 2024–January 30, 2024), and Sector 75 (January 30, 2024–February 26, 2024). 

The aforementioned observations provide a comprehensive dataset up to the present date, and the corresponding light curves and periodogram are shown in 
Figure \ref{fig8}. A very short period of $\sim$ 6 days is obtained from the TESS data (see bottom right panel). The phase-folded light curve and its moving median help us illustrate the repetitive structure of the combined light curve (see the bottom left panel). We were unable to get a long-term period of several hundred days, spread over several sectors, even though the TESS mission observed I19520 for a considerable amount of time ($\sim$ 1000 days) with gaps between sectors. This could be related to the fact that CBVs, employed to eliminate systematic effects during the post-processing of each TESS sector, may have inadvertently eliminated genuine long-term periods from the data.
We further performed an individual exploration of the periods identified in each TESS sector (cf. Figure \ref{fig8n}). Our analysis shows that the periods detected in individual sectors are consistent with the results obtained from the multi-sector analysis. While sectors 74 and 75 display the most apparent periodic variations visually, the periodic signals are also present, though weaker, in other sectors. This confirms that the detected period is robust across multiple TESS observations, with minor variations likely due to differences in noise levels, data coverage, or intrinsic variability.

\begin{figure*}
\centering
\includegraphics[width=1\textwidth]{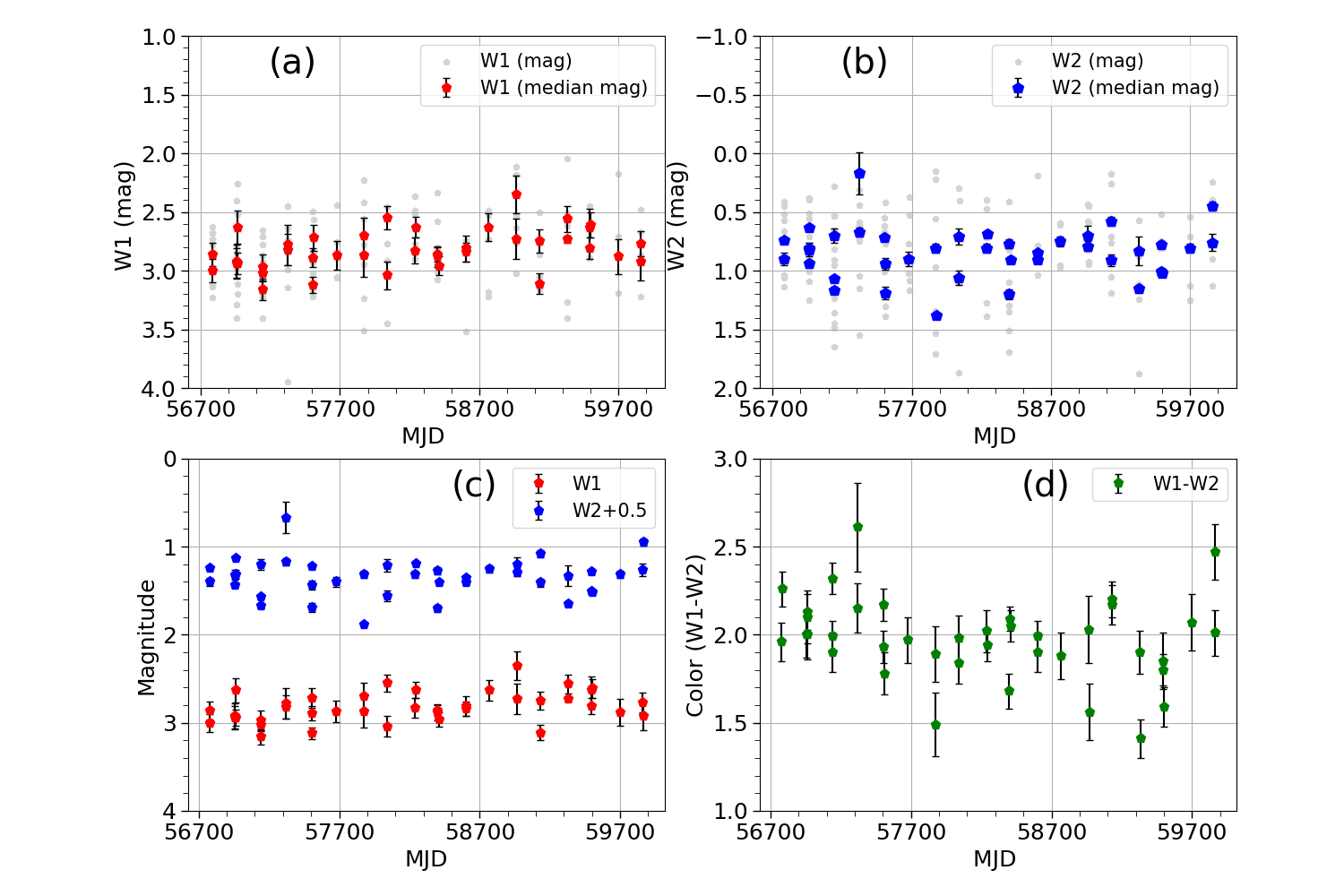}
\caption{\label{fig9} Light curve of I19520 in NEOWISE W1 (a) and W2 (b) bands. The grey dots in both plots show all the data points, while the colour dots are the median magnitudes, as discussed in Section \ref{sec34}. (c) Median light curve of the source in W1 (red) and W2 (blue) bands. (d) Colour (W1-W2) of the source plotted against the MJD.}
\end{figure*}

\js{\begin{figure*}
\centering
\includegraphics[width=0.4\textwidth]{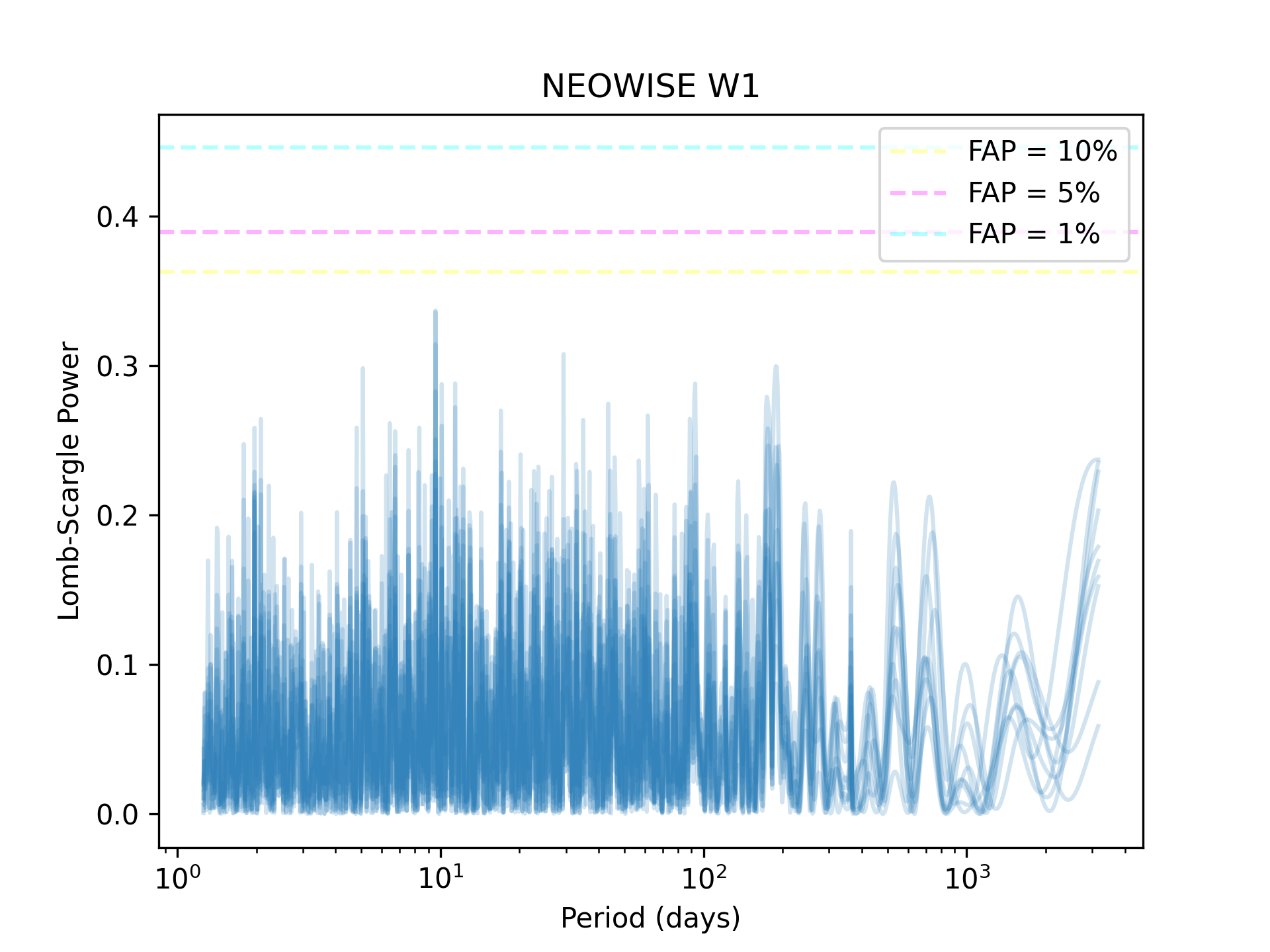}
\includegraphics[width=0.4\textwidth]{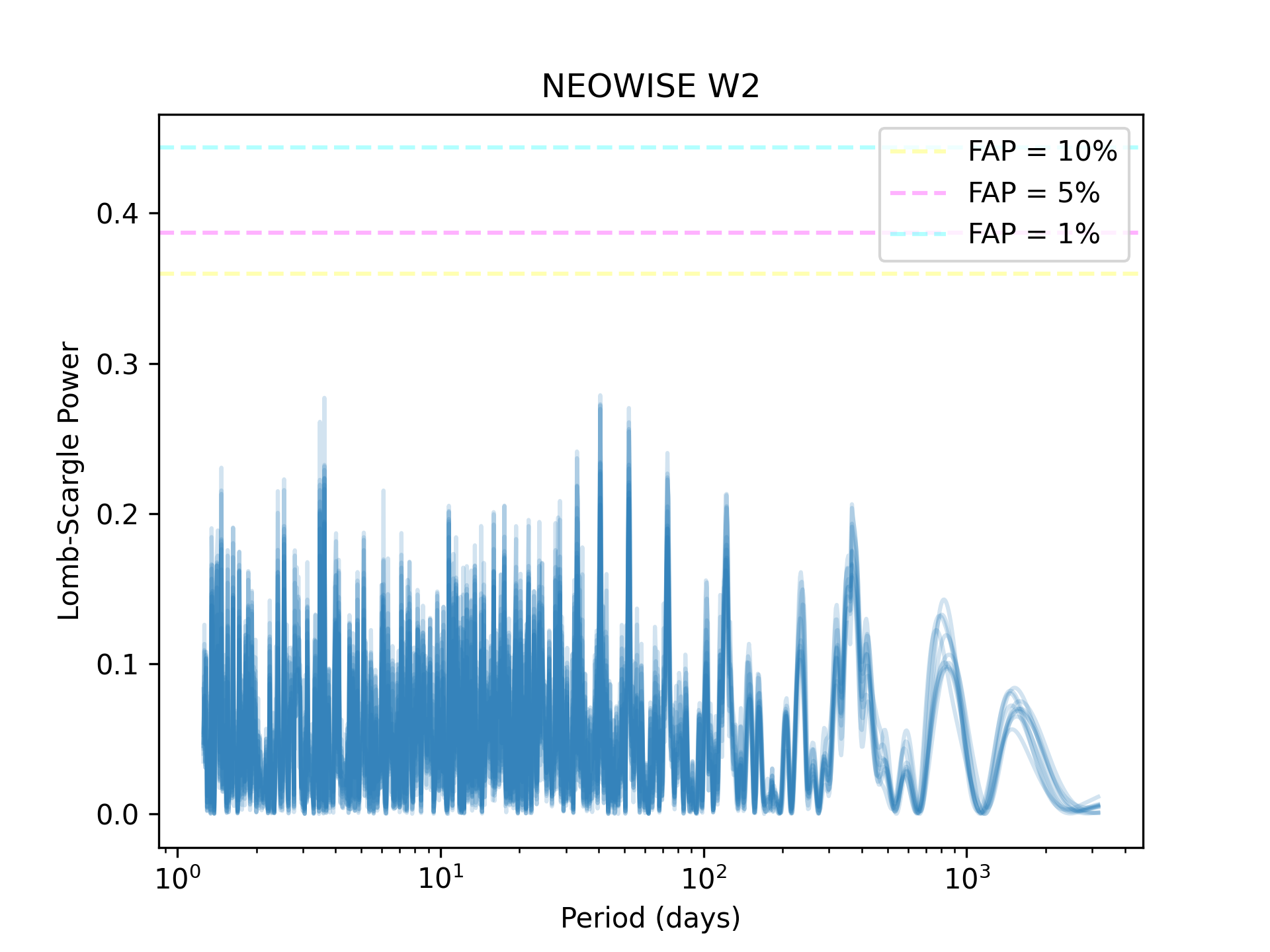}
\caption{\label{wi} Lomb-Scargle Periodogram for NEOWISE W1 band (left panel) and W2 band (right panel) with 1\%, 5\%, and
10\% FAP levels. The thresholds in both periodograms indicate the absence of a significant periodic signal in these data sets.}
\end{figure*}}

\subsection{NEOWISE light curves and period analysis} \label{sec34}
We also plotted in Figure \ref{fig9} the light curve of the source in the W1 and W2 bands using the NEOWISE data. For the NEOWISE data set, the characteristic transition from non-saturated to saturated detections is not very sharp and occurs near W1 = 8.0 mag and W2 = 7.0 mag. Photometry has been performed for all NEOWISE detections by fitting point-spread-function (PSF) templates to the source images normally over a radius of 1.25 times the FWHM of the PSF. For very bright sources that saturate the detectors, the fitting radius is enlarged (twice the radius of the saturated region), and the fit is performed on the non-saturated pixels in the wings of the source profiles. We exclude measurements W1 $<$ 2 mag and W2 $<$ 0 mag from our analysis since they are deemed useless. The explanation supplement to the NEOWISE data release provides a thorough explanation. Additionally, for the bright sources, both band magnitudes need to be corrected for colour-dependent saturation effects. However, the correction values for magnitudes brighter than W1 $\sim$ 2.7 mag and W2 $\sim$ 2.5 mag are not considered reliable\footnote{\url{https://wise2.ipac.caltech.edu/docs/release/neowise/expsup/index.html}} and therefore, since they are not given in the explanatory supplement to the NEOWISE data release (cf. Table 2), we could not correct our source magnitudes (mean W2 mag around 1 mag) for the colour-dependent saturation effects. Therefore, the results given below should be regarded with caution.

The survey provides multiple magnitude measurements corresponding to a given MJD, which shows considerable scatter. It has been a common approach to get a median or mean magnitude corresponding to each MJD for further analysis of the light curve \citep{ghosh23,uchi19,uchi22}. We took the median of the magnitude and the error of all corresponding measurements for a given MJD. The median magnitude is overplotted in the light curves of Fig. \ref{fig9} with coloured dots. We also compared the median light curves in the bottom left panel, while the bottom right panel shows the variation of the colour (W1-W2) with the MJD. The source is found to be variable in NEOWISE data, with a variation of $\sim$ 1 mag in the W1 and W2 bands.
We further performed LSP analysis for the NEOWISE data to estimate any potential periods in the light curves. The power spectra of the source in the W1 and W2 bands, along with the FAP levels of 1\%, 5\%, and 10\%, are shown in Figure \ref{wi}. The figure clearly shows that there is no significant period present in any of these bands.


\begin{figure}
\centering
\includegraphics[width=.5\textwidth]{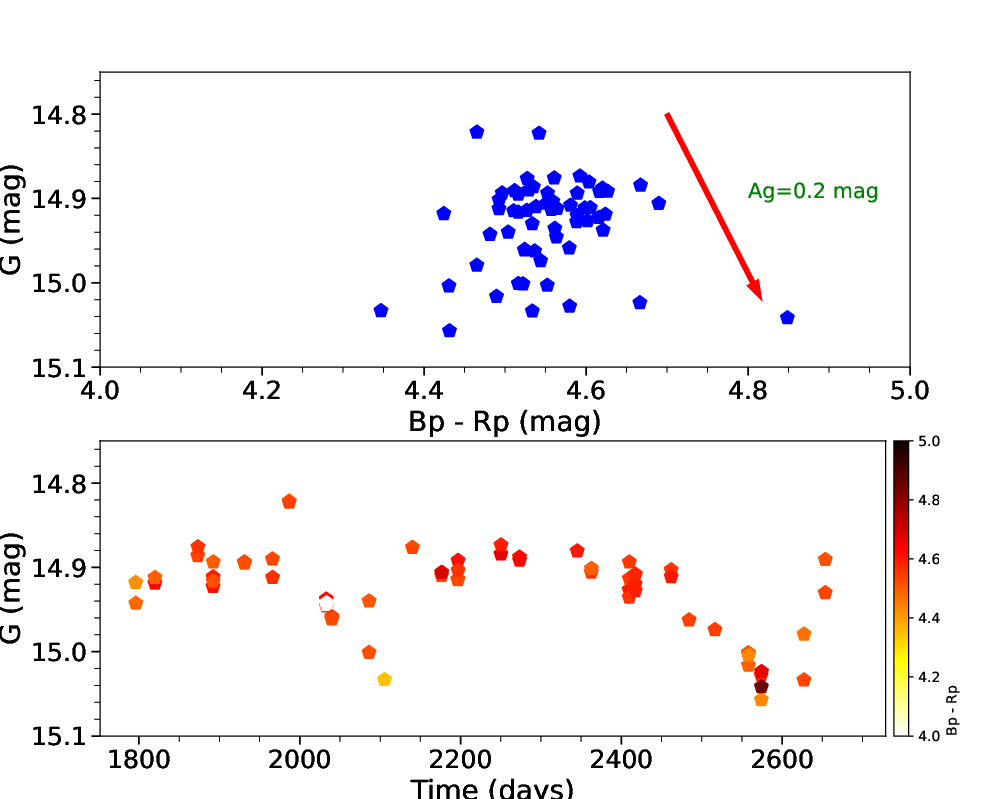}
\caption{\label{fig10} (Top) Gaia G vs. (Bp-Rp) CMD of the source. (Bottom) Colour (Bp-Rp) coded light curve of the source in the Gaia G band. The arrow in the top panel shows the extinction vector corresponding to 0.2 mag in the G band.}
\end{figure}

\begin{figure}
\centering
\includegraphics[width=.5\textwidth]{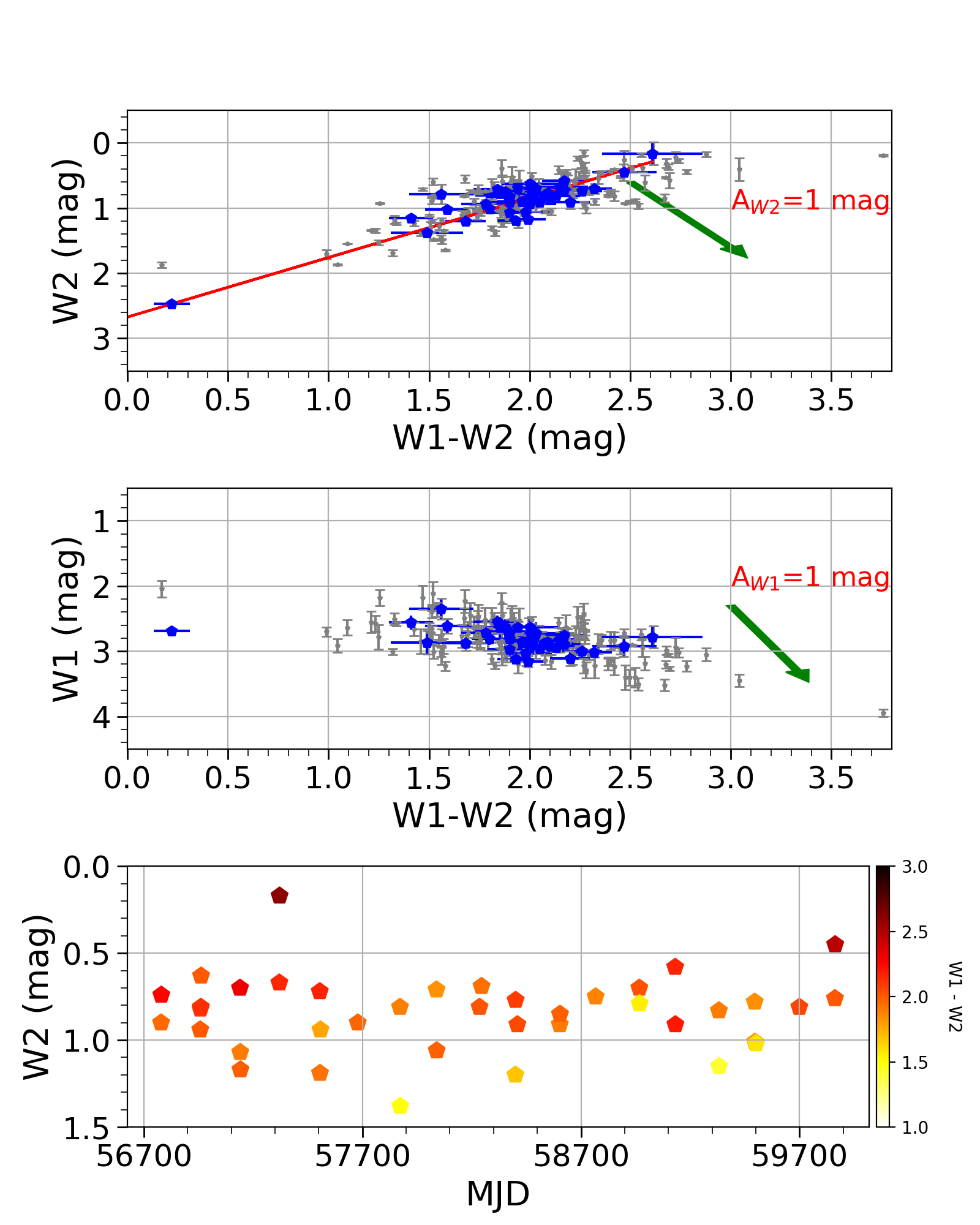}
\caption{\label{fig11}  The top panel shows the W2 vs (W1-W2) CMD of the source, the middle panel shows the W1 vs (W1-W2) CMD, and the bottom panel shows the colour (W1-W2) coded W2 band light curve.}
\end{figure}

\subsection{Colour variations in I19520} \label{sec35}

YSOs are usually found to show colour variations along with variations in the flux. The different trends and the magnitude of colour variations can be produced by different types of physical mechanisms powering the variability \citep{Gunther14}. Hence, colour variations have proven to be an important tool for identifying and differentiating among the various possible mechanisms \citep{carp01,Gunther14,conteras16}. We investigated the colour variation in I19520 using Gaia and NEOWISE data sets. We plotted the colour-magnitude diagram (CMD) of the source using Gaia data sets (G vs. Bp-Rp) in the top panel of Figure \ref{fig10}. The arrow shows the reddening vector of 0.2 mag in the G band as per the relation given by \citet{wang2019}.  The Gaia data shows a negative slope in the CMD opposite to the direction of the reddening vector which signifies that the source is getting bluer while getting fainter (BF). The bottom panel shows the light curve of the source in the G band with each point marked with its corresponding colour. This figure also reflects a similar trend as the top panel.

We also plotted the CMD of the source using the NEOWISE data sets.
The W2 vs (W1-W2) CMD of the source is shown in Figure \ref{fig11}. The grey dots in the background show all data points with the corresponding errors, while the blue dots depict the median data points as discussed in Section \ref{sec34}. The figure clearly shows a negative slope in the CMD (BF trend) which seems orthogonal to the reddening vector \citep[shown with the green arrow,][]{wang2019} a bit similar to the optical CMD.  In the W1 vs (W1-W2) CMD, as shown in the figure, the magnitude (blue dots) seems to be almost constant as the colour varies. We have also shown the light curve of the source in the W2 band with each point marked with the corresponding (W1-W2) colour. 
As both axes of the  W2 vs W1-W2 CMD contain W2, 
an artificial $y=-x$ linear sequence resulting in an artificial slope of $\pm$45$^{\circ}$ can be apparent in the CMD \citep{Gunther14}.
To determine the slope of W2 vs W1-W2 CMD we used the orthogonal distance regression (ODR) method. The ODR method is quite useful and more accurate than the normal regression method when both axes are prone to errors. This method is used by several authors studying MYSOs variability using NEOWISE data \citep[e.g.,][]{uchi19,uchi22}. Since we are interested in the long-period variation of the source, we considered median data points for the fitting. We obtained the equation of the ODR fitting as  $y=(-0.91\pm0.04)x+(2.67\pm0.08)$, and the fitted line is shown in Figure \ref{fig11}. The obtained slope of the CMD within errors is very close to $-45^{\circ}$.
The ODR analysis thus indicates that it is not possible at this time to verify the observed BF trend in optical data using the NEOWISE data sets. Nevertheless, the colour variation does disfavour the bluer while getting brighter (BB) trend.


\subsection{Tentative physical parameters of I19520 from Inayoshi's relations} \label{sec36}
We derived some of the physical parameters of I19520 based on the predictions of \citet{inayoshi2013}. The PL relation given by \citet{inayoshi2013} is:
\begin{equation}
\log (L/~\lsun) = 4.62 + 0.98\log(P/100~{\rm days}).
\label{eq:P-L}
\end{equation}
The authors additionally derived equations to determine parameters such as mass, radius, and accretion rate of the bloated star considering the spherical accretion scenario. The dependence of these three parameters on the period of the source is expressed as \citep{inayoshi2013}:

\begin{equation}
M_{\ast} = 17.5~\msun \left(\frac{P}{100~{\rm days}}\right)^{0.30},
\label{eq:MP}
\end{equation}
\begin{equation}
R_{\ast} = 350~\rsun \left(\frac{P}{100~{\rm days}}\right)^{0.62},
\label{eq:RP}
\end{equation}
\begin{equation}
\dot{M}_{\ast} = 3.1\times 10^{-3}~\msunyr 
\left(\frac{P}{100~{\rm days}}\right)^{0.73}.
\label{eq:MdotP}
\end{equation}

Since the period of 240--270 days for I19520 ($\sim10^5\,\lsun$) is very consistent with the PL relation considering a spherical accretion scenario, we derived these parameters using the above equations using a period of 270$\pm$50 days (the optimum value).
We obtained $M_{\ast}\approx23.6\pm1.1\,\msun$, 
$R_{\ast}\approx650\pm70\,\rsun$, and 
$\mdot\approx(6.4\pm0.8)\times10^{-3}\,\msunyr$.
However, the 240--270 days period could be an harmonic (cf. Sec. \ref{sec32}) of the more robust period determined with Gaia, of 440--460 days, but using a period of 460 days in equations (2)--(4) yields $M_{\ast}\sim 28\,\msun$, $R_{\ast}\sim900\,\rsun$, and 
$\mdot\sim 9\times10^{-3}\,\msunyr$, which are unrealisticly high.
This is further discussed below.


\section{Discussion} \label{sec4}
 
\subsection{More realistic physical parameters for I19520} \label{sec40}

By using \citet{inayoshi2013} equations outlined above, a mass of $\sim 24$--28$\,\msun$, a radius of $\sim 650$--900$\,\rsun$ and a mass accretion rate of $\sim (6$--$9)\times10^{-3}\,\msunyr$ were obtained for I19520.
The derived range for the mass is in good agreement with the mass estimated by \citet{Palau2013} using the SED analysis (26--33 $\msun$).
However, the SED analysis of \citet{Palau2013} did not consider that the star is bloated and, additionally, equations (1) to (4) are highly depending on the geometry of accretion \citep{inayoshi2013}. In particular, the derived mass accretion rate close to 10$^{-2}$~$\msunyr$ is too high compared to the expected value for high-mass protostars. Mass accretion rates can be guessed from estimates of mass infall rates on clump scales, as the first are supposed to be about one order of magnitude smaller than the second \citep[e.g.,][]{DunhamVorobyov2012, Mottram2013}. Mass-infall rates have been measured to be of the order of 10$^{-4}$--$10^{-2}$\,$\msunyr$ for $10^4$--$10^5$~$\lsun$ massive protostars, according to the correlation between mass infall rate and bolometric luminosity reported in Figure 5 of \citet{Palau2024}, and many other individual measurements \citep[e.g.,][]{Olguin2023, Xu2023_SDC335, Mookerjea2023, Yang2023_Infall, He2023_Infall, Xu_2023_Infall}. On the other hand, mass accretion rates have been found to be in the range 10$^{-6}$--$10^{-3}$\,$\msunyr$ \citep[e.g.,][]{Sanhueza2021, Beuther2023, Cesaroni2025}, about one order of magnitude lower than their corresponding mass infall rates on clump scales. Regarding I19520, \citet{Palau2013} estimate the mass infall rate to be around $3\times10^{-4}$~$\msunyr$. This estimate has an uncertainty of about a factor of 2--3 from the $^{13}$CO abundance used in \citet{Palau2013}, for which a specific $^{12}$C/$^{13}$C isotope ratio of 1/89 was used while this ratio is known to vary in different galactic environments \citep[e.g.,][]{Giannetti2014_isotopicratios, Sun2024_isotopicratios}.
Because of the reasons given above, it seems reasonable to consider that the accretion rate for I19520 cannot be much larger than 10$^{-3}$~$\msunyr$, well below the value obtained from equation (4) considering a period of 270--460 days.

A possibility to reconcile the high value for the accretion rate obtained from equation (4) and the expected upper limit for I19520 of 10$^{-3}$~$\msunyr$ is to consider that the accretion is not spherical but through a disc.
As predicted by \citet{inayoshi2013}, the period of pulsations for the disc accretion case should be higher than that predicted by the PL relation. In fact, the period of 460$\pm$80 days seems to be well consistent with the disc accretion scenario (cf. Figure \ref{fig12}). 
If the accretion in I19520 is proceeding through a disc rather than spherically, the relations given in equations (2) to (4) cannot be used.

In order to have an estimate of the radius of I19520 for the disc accretion case, we used its bolometric luminosity and effective temperature and applied the Stefan–Boltzmann (SB) relation. We adopted a bolometric luminosity of $10^{5}$ $\lsun$ (see section \ref{sec1}) and a temperature of $\sim17000$ K, typical of the B3 spectral type \citep{pangia73,Fitzpatrick2007}, which seems most plausible for I19520 (S\'anchez Contreras, priv. communication). In this case, the SB relation yields a radius of $\sim36\,\rsun$, much smaller than the value estimated using equation (3).  Also, according to the Hertzsprung–Russell (HR) diagram from \citet{hoskawa2010} (Figure 14), for an effective temperature of log(Teff) = 4.25 (approximately 17,000 K), the radius is around 40 R$\odot$, and the mass is estimated to be around $\sim15$~M$\odot$. 
Additionally, if we consider an accretion rate of 10$^{-3}$~$\msunyr$ (see the justification above), with the radius of 36\,$\rsun$, the star seems to be well into a bloating phase as per the evolutionary track for the disc accretion scenario presented in Figure 12 of \citet{hoskawa2010}. From the figure of that paper it is clear that the radius is significantly reduced in the disc accretion case compared to the spherical accretion case. Thus, the disc accretion scenario seems to be a more plausible scenario, which is also well consistent with the fact that the source is known to be driving a collimated outflow.

It is also important to keep in mind that the effective temperatures assumed in \citet{inayoshi2013} are taken from \citet{Hosokawa2009} tracks as input to calculate the corresponding relations given in equations (2) to (4). However, the \citet{Hosokawa2009} tracks do not include the most likely effective temperature of I19520 of around 17000 K (see Fig. 6-bottom of \citealt{Hosokawa2009}). Further theoretical predictions for the disc accretion case and for larger effective temperatures would be extremely useful for estimating the physical parameters of bloated MYSOs using their period.

\subsection{Bloated MYSOs and stellar pulsations} \label{sec41}

A high accretion rate is one of the factors distinguishing the high-mass star formation from its low-mass counterparts. Since massive stars have short Kelvin–Helmholtz (KH) timescales and they evolve very fast (roughly 10$^{5}$ years), that puts a lower limit in the accretion rate ($\sim$ 10$^{-4}$ $\msunyr$ for a 30 $\msun$ star; \citealt{yb08}). 
It has also been suggested that such high accretion rates are necessary for the accretion flow to withstand the intense radiation pressure from the developing massive star \citep[e.g.,][] {larson71,wolf87}.
Several observational works also report mass accretion rates $\gtrsim10^{-4}$ $\msunyr$ for high-mass protostellar objects (see references in Sec.~\ref{sec40}). 

A number of theoretical works studying the protostellar evolution with a high accretion rate suggest that stars swell up while accreting \citep{pala92,yb08,Hosokawa2009,hoskawa2010,inayoshi2013,kuiper13,hle2016}. The radius of the star can reach up to 30--400 $\rsun$ for a 10 $\msun$ star \citep{Hosokawa2009}. This kind of swelling is found to be a robust feature of accretion at a high rate, independent of the adopted initial models and for both spherical and disc accretion \citep{Hosokawa2009, hoskawa2010,kuiper13}. The swelling is considered as one of the four evolutionary phases of protostellar accretion, and these four phases are summarized below for the disc accretion case. The main driver of the evolution in these four phases is the gradual decrease of the opacity inside the stellar interior. As the star gains mass from accretion, the temperature increases in the stellar interior. Since the opacity in stellar interiors, $\kappa$, follows the Kramer's law, by which $\kappa \propto \rho\,T^{-3.5}$ \citep[e.g.,][with $\rho$ being the density]{hoskawa2010}, an increase in temperature implies a strong decrease in opacity.

\noindent
\textbf{Phase 1}: 
With faster mass accretion, the accreting gas has a greater specific entropy, which raises the average entropy in the stellar interior. The initial phase is known as the convective phase, during which the opacity is high enough to allow efficient accumulation of entropy within the star. \\
\textbf{Phase 2}: 
As the opacity decreases, the gradual outward transport of entropy on a thin layer of the stellar surface causes it to swell up (the swelling phase). \\
\textbf{Phase 3}:
After the opacity falls sufficiently so that the star starts to lose heat efficiently via radiation, it starts to contract to maintain virial equilibrium. \\
\textbf{Phase 4}:
Due to the contraction, the temperature and density start to rise, and as the hydrogen burning starts, the star enters the main-sequence accretion phase.

Although the sequence of evolution described above is for the disc accretion case, spherical accretion also goes through the swelling phase, with some variation in the other phases \citep[cf.][]{hoskawa2010}. The variation in the stellar radius for different accretion rates as the accretion proceeds (both disc and spherical accretion) is shown in Fig.~12 of \citet{hoskawa2010}. 

The linear stability study by \citet{inayoshi2013} on the protostellar models of \citet{Hosokawa2009} suggests that the star becomes pulsationally unstable during the bloated phase. The authors noted the $\kappa$ mechanism \citep{cox63,cox80} in the  He$^{+}$ ionization layer as the cause of the instability. As the star moves into the KH contraction phase, the He$^{+}$ ionization ceases to exist, as do the pulsations.
In their study, only protostellar models with accretion rates > 10$^{-3}$ $\msunyr$ develop stellar pulsations. Although the protostellar models for lower accretion rates became unstable, they fail to develop stellar pulsations due to the large time taken by perturbations to grow (see Figure 2 of \citealt{inayoshi2013}). 

The PL relationship of Eq.~\ref{eq:P-L} arising from the $\kappa$ mechanism in the He$^+$ ionization layer, as derived by \citet{inayoshi2013}, is shown in Figure \ref{fig12}. The relation is derived using the protostellar models of spherical accretion (for different accretion rates), but the authors show that the protostar also becomes pulsationally unstable in the disc accretion case. However, the period in the disc accretion case is suggested to be multiple times greater than that in the spherical case.


The authors also explain the periodic variability in the 6.7 GHz methanol masers sources with the PL relation. In Figure \ref{fig12}, maser sources with periodic variations are also shown; the empty squares specifically represent the sources devoid of free-free emission representing bloated star candidates. \citet{inayoshi2013} argue that the maser emission arises in the accretion disc and is radiatively pumped by the protostar. Thus, the pulsations in the protostar itself could cause flux variations in maser sources (see \citealt{inayoshi2013} for further details). Interestingly, correlations between the protostellar flux variations and the maser variability have been observed for some of the sources \citep{olech20,uchi22}. One crucial point to notice is that the PL relationship holds seemingly well for the sources lacking free-free emission, while sources with important free-free emission show a significant departure from it, most likely because they are not bloated any more. Another point is that the period of the maser sources is higher than predicted by the PL relation. This is explained as a deviation from the spherical accretion. 
It is important to note that the pulsation model is the only model that can explain the maser variability in sources with no free-free emission, as alternate models require the presence of an Ultra-Compact HII Region (UCHII) surrounding the star \citep{vd11}. Thus the PL relation proposed by \citet{inayoshi2013} has only been contrasted observationally with methanol maser variability in previous works. 

\subsection{I19520: An optically visible accreting MYSO} \label{sec42}

The above discussion opens a path for bloated stars to be tested by the pulsation model of \citet{inayoshi2013}. In the past, some MYSOs had been classified as bloated star candidates \citep{morino98,linz09,och2011}, but none of them have been tested for the pulsational model through optical/infrared variability. 
I19520 is the first bloated star candidate where such a test has been carried out.

As mentioned in Sec.~\ref{sec1}, I19520 has a strong compact millimetre source associated with the optical source, and no centimetre detection \citep{Palau2013}.
The association of the bright optical source with the submillimetre clump has been a matter of debate (see \citealt{Palau2013} for more details).
The reprocessing of the shorter-wavelength radiation into longer-wavelength (MIR to FIR) due to the presence of dust in the protostellar envelope could inhibit the star from being visible at optical wavelengths. However, if the outflow cavity is partially aligned along the line of sight, the source could be visible at optical wavelengths \citep{2007prpl.conf..117W,1998AJ....115.2491K}. 
\citet{Palau2013} confirmed a collimated outflow associated with the source, indicating that the envelope may contain cavities and holes. The best-fit model in the SED analysis has a cavity with a full opening angle of 62$^{\circ}$, and the axis of the cavity is inclined to the plane of sky by 59$^{\circ}$, so it is not surprising that some fraction of the optical light from the star reaches us.
Additionally, the positional agreement between the optical/IR and submillimere source is also suggested by \citet{Palau2013}.
Thus, I19520 proves to be a MYSO, still in an accreting phase, although visible at optical wavelengths. Accreting MYSOs detected in the optical are rare but not inexistent. A recent study by \citet{mck2024} reports an optically visible MYSO accreting through a Keplerian disc in the Large Magellanic Cloud.

\subsection{I19520: A long-period variable YSO} \label{sec43}
YSOs are known to show photometric variability, observable from almost X-ray to optical wavelengths. Different phenomena, such as spots in the photosphere, accretion, variable extinction, eclipsing binary or a close companion, pulsations, etc., are known to cause variations in the brightness of YSOs (\citealt{conteras16}, and references therein).
Depending on the temporal variation of the brightness and the magnitude of the variation, they are further subcategorized into different subclasses.
Some of the most popular subclasses are short-period variables (SPV), long-period variables (LPV), dipper and faders, as well as eruptive variables \citep{conteras16,kumar16,texera18}. Among them, the SPVs and LPVs are the ones showing periodic variations in their brightness, typically for < 100 days (SPV) and >100 days (LPV), respectively. 

Based on the theoretical PL relation, I19520 was expected to fall in the subcategory of a LPV, as indeed found in this work.
The LPV population is usually found to be important in samples of variable YSOs (from here on, we will refer to them as YSO-LPVs). For example, in the sample by \citet{conteras16}, $\sim$ 15 \% of the sources are classified as YSO-LPVs. In the higher mass regime, $\sim$ 32 \% of MYSOs in the sample of \citet{texera18} are YSO-LPVs. 
Interestingly, the YSO-LPV category is particularly vulnerable to contamination by non-YSO sources, including dusty AGB stars and Mira type variables.
Particularly, dusty AGB stars with colours similar to those of YSOs and with a periodic variation of typically 400 <P <2500 days \citep{wlock2008} could be an important source of contamination.
However, a key differentiator in the light curve of YSO-LPVs and AGB stars is the presence of short-scale patterns or scatters. In the light curve of YSO-LPVs, small scatters are superimposed on the long-period variation, while AGB stars usually show more smooth sinusoidal variations. 
The short-scale scatters in the light curve of YSO-LPVs are potentially caused by hot and cold spots, small variations in accretion rate, or extinction. The scatter in the I19520 light curve (cf. Fig. \ref{fig2}) is thus more consistent with the young star nature than the AGB nature. This is also consistent with the millimetre continuum and gas emission reported by \citet{Palau2013}, suggesting an envelope mass for I19520 of about 140~$\msun$. Hence, the possibility of an AGB nature for I19520 is completely ruled out.

\subsection{Nature of variability in the MYSO I19520 } \label{sec44}

\subsubsection{Previous variability studies in MYSOs}

MYSOs have been the subject of variability studies in some of the past and recent research works \citep[e.g.,][]{kumar16,texera18,uchi19,uchi22}. In these studies, MYSOs are found to show photometric variability at different timescales and of different types as their lower mass counterparts. 
However, these studies are less frequent as compared to low-mass studies due to the scarcity of MYSOs. 
One of the largest samples of variable MYSOs to date has been presented in \citet{texera18}. 
They used multi-epoch NIR observations of the Vista Variables in the Via Lactea (VVV) survey to study the photometric variability in a sample of 718 MYSOs. A significant part of their sample is comprised of Spitzer-extended green objects (EGOs), known to be associated with outflows \citep[e.g.,][]{Cyganowski2008_EGOs}, while others are bright 24 $\mu$m sources. 
In that study, 91\% of the targets classified as EGOs are variable ($\Delta$Ks >0.14 mag) in the NIR Ks band. Based on that, the authors suggest that the variability in MYSOs is closely linked to accretion and outflow processes.

A smaller sample of 13 MYSOs (associated with the peak of ATLASGAL clumps) is found to be variable ($\Delta$Ks > 1 mag) by \citet{kumar16}. 
The nature of the variability is suggested to be the same phenomena producing variation in maser sources, i.e., 
the change in IR luminosity due to episodic accretion, in line with \citet{texera18}.
A more recent study by \citet{uchi19} report MIR variability in five MYSOs (out of 650), including one periodic source. The variability is suggested to be caused by either the variation in the extinction along the line of sight or the change in mass accretion rate. 

The pulsational model has been invoked in past studies \citep{texera18,uchi19} as one of the plausible mechanisms behind the MYSO variability. \citet{uchi19} discussed that the observed periodic variability in the MYSO G335.9960–00.8532 could be explained by the predicted stellar pulsational model. However, the derived period is not consistent with the bolometric luminosity of the source, considering the PL relation. The authors argued that if the far-side distance of the source is considered, it would be consistent with the PL relation by a factor of three. Similarly, \citet{texera18} suggest that eruptive MYSOs most likely enter a bloated phase inducing pulsations as a consequence of the high accretion rate, before re-adjusting and contracting again.

\begin{figure}
\centering
\includegraphics[width=.5\textwidth]{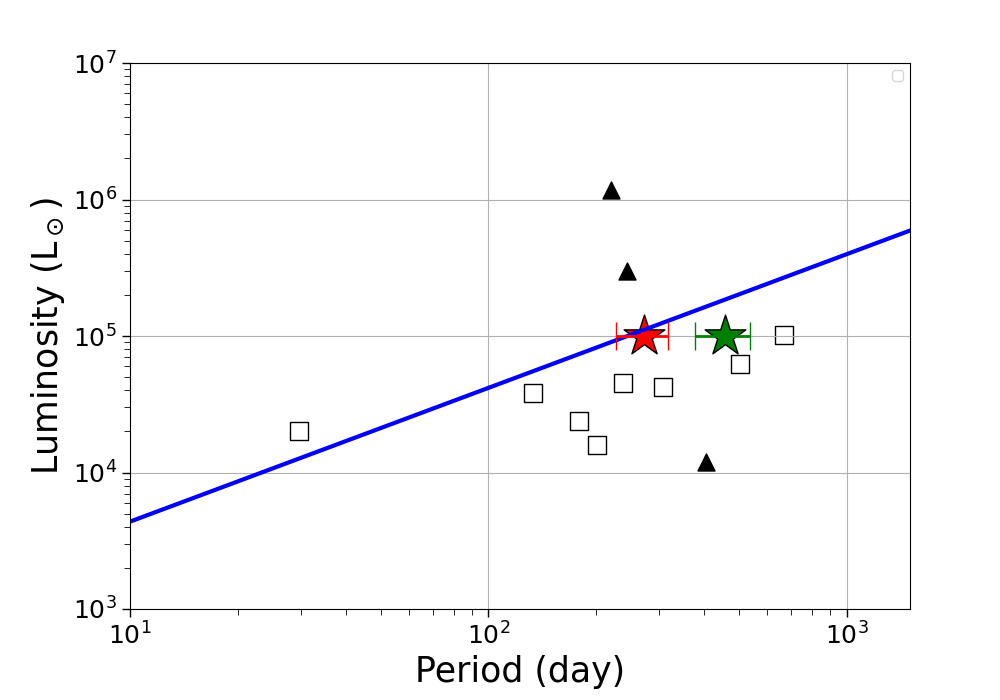}
\caption{\label{fig12} 
Luminosity vs. Period relation from \citet{inayoshi2013} corresponding to the spherical accretion case (thick blue line). The empty squares in the figure represent the maser sources lacking free-free emission being bloated star candidates, while the triangles represent sources with free-free emission. The position of I19520 is marked with the red star symbol, considering a period of 270 days (with an error bar of 50 days), and with a green star symbol, corresponding to a period of 460 days (with an error bar of 80 days).}
\end{figure}

\subsubsection{Variability in I19520: consistent with pulsations due to a bloated phase}

In the present analysis, I19520 is found to be variable in the optical and infrared bands, clearly showing long periodic variations in the optical data. We obtain a short period of about 270 days ($\sim$ 270$\pm$40 days and $\sim$ 270$\pm$50 days in the TJO Rc and Ic bands, respectively). A similar short period (within uncertainties) is found using the Gaia data ($\sim$ 250$\pm$20 days and $\sim$ 240$\pm$20 days from the Gaia G and Rp bands, respectively). The short period (shown with a red star symbol in Figure \ref{fig12}) is in good agreement with that predicted for I19520 by the PL relation for the spherical accretion phase -- 244 days for I19520's bolometric luminosity of $\sim10^{5}\,\lsun$. 

A second longer period of about 450 days is also found from fitting the Gaia data ($\sim$ 460 $\pm$ 80 days (G band) and $\sim$ 440 $\pm$ 70 days (Rp band)) that is in better agreement with the theoretical prediction for the disc accretion case (\citealt{inayoshi2013}, cf. Section \ref{sec41}), in which the period can be an integer multiple of the period derived for the spherical accretion case. It is important to mention that in the Gaia light curve analysis (cf. Section \ref{sec32}), the longer period is found to be more reliable than the period consistent with the spherical accretion scenario (of the order of 244 days). Thus, we consider the longer period (shown with a green star symbol in Figure \ref{fig12}), which favours the disc accretion scenario, to be more robust and is also consistent with the observed variability in maser sources. The disc accretion scenario is also fully consistent with the fact that I19520 is known to be the driving source of a collimated molecular outflow \citep{Palau2013}. The obtained physical parameters of the source (as discussed in Section \ref{sec40}) also favour this conclusion. Additionally, the intrinsic change in the source flux, such as that due to stellar pulsations, could be reflected as a negative slope in the CMD \citep{Gunther14,uchi19}, and the observed colour trend in I19520 (cf. Section \ref{sec35}) is consistent with that. Thus, the pulsational model could well explain the observed variability in I19520.



However, 
alternative scenarios cannot be entirely ruled out, and
the feasibility of other plausible interpretations for the observed characteristics of the light curve of I19520 is discussed in the following subsection.

\subsubsection{Variability in I19520: other possible mechanisms additional to pulsation}

Other mechanisms, in addition to pulsation due to a bloated phase, could be related to the variability in I19520. The most important ones are summarized and discussed below. 

\paragraph*{Modulation by stellar rotation:} Stellar rotation can cause periodic variation in the light curve by modulating the cool and hot spots, but the variation timescale is typically $<15$ days \citep{conteras16}. A variability of >100 days, as noticed in I19520, is very unlikely to be produced by stellar rotation \citep{uchi19}. Additionally, both cool and hot spots tend to produce either a positive slope in the CMD or no colour variation \citep{carp01}. In contrast, the observed colour trend at optical wavelengths, which is more sensitive to spots and extinction events \citep{Gunther14}, indicates the opposite (cf. Section \ref{sec34}). Nonetheless, the TESS data yielded a smaller period of $\sim$ 6 days (see Section \ref{sec33}), which could be caused by stellar rotation.

\paragraph*{Extinction variation along the line-of-sight:} Variation of extinction along the line-of-sight is another primary source of variability in YSOs \citep{carp01,Gunther14,conteras16}. Asymmetries in an inner disc or the circumstellar environment could cause periodic variability due to the Keplerian motion of the disc. Additionally, there could be variations of extinction produced by variations in the local properties of the natal cloud. In this last case, the variability is not necessarily periodic. The change of extinction in the line-of-sight tends to show a positive slope in the CMD, i.e., the source gets bluer while getting brighter (BB) \citep{carp01,Gunther14}. The colour trend observed in I19520 (the negative slope in the CMD) is not consistent with the extinction variation along the line-of-sight and is therefore ruled out.

\paragraph*{Interaction with a binary:}
The interaction with a binary, which is very common in high-mass star formation \citep[e.g.,][]{chini2012}, can cause a periodic change in the accretion rate, a phenomenon which is termed `pulsed accretion' \citep{araya2010, Muze2013}. This in turn will cause a periodic change in the luminosity of the source, producing periodic variability. Since varying accretion could reflect as a negative slope in the CMD \citep{carp01,Gunther14,uchi19}, which is consistent with the observed colour trend in I19520 (cf. Section \ref{sec35}), this possibility cannot be ruled out.

\paragraph*{Gravitational instability:} Gas densities needed to provide high accretion rates, required for massive star formation, could induce gravitational instabilities in the accretion disc \citep{meyer17,meyer19,oliva20}. Fragments produced by the gravitational instability \citep{oliva20} are usually considered responsible for episodic accretion events driving outbursts in MYSOs \citep{meyer17,meyer19}. However, this could also occasionally cause periodic changes in the accretion rate \citep{matu2017,uchi19}, in a similar fashion as changes produced by binary interaction. Thus, in this case, a negative slope in the CMD is also expected, making this possibility plausible. \\

In summary, changes in the accretion rate (produced either by a binary or gravitational instabilities) could potentially explain the light curve characteristics of I19520, in addition to the bloated star pulsations. To examine the scenario of the accretion rate change, a detailed spectroscopic time series analysis in the optical and NIR bands is required to monitor and examine the variations in accretion tracers such as hydrogen recombination lines \citep[e.g.,][]{basri97,jsen2007}.
\\

Consequently, a multi-band variability study spanning different wavelength regimes, including both photometry and spectroscopy, is essential to weigh the different possible mechanisms present in I19520. This work demonstrates the relevance of variability in bloated stars and pushes to search for further candidates and to build samples of these objects, so that variability studies can be performed from a statistical point of view.

\section{Summary and conclusions} \label{sec5}

We have used optical and infrared time series data to perform a variability study on a candidate bloated MYSO, I19520. We tested the PL relation proposed in the literature to investigate the bloated star hypothesis for this source. Our study leads us to the following conclusions:

\begin{itemize}
\item
The source is found to be variable in optical (TJO, Gaia, and TESS) and infrared (NEOWISE) bands, showing clear periodic variations in optical bands. No significant period is evident in the NEOWISE bands.

\item
The optical light curve of the source is similar to an LPV YSO with small-scale scatters or periods superimposed on a long period. A periodogram analysis of the Gaia data yields a strong period of $\sim$ 460$\pm$80 days (in the G band) and $\sim$ 440$\pm$70 days (in the Rp band). The Gaia data also reveal a second period of $\sim$ 250$\pm$30 days (in the G band) and $\sim$ 240$\pm$20 days (in the Rp band), which are consistent with the periods found in the TJO data of $\sim$ 270$\pm$40 days (in the Rc band) and $\sim$ 270$\pm$50 days (in the Ic band). A very short period of $\sim$ 6 days is obtained from the TESS data. 

\item
The short period of 240--270 days (with an error of 20--50 days) fits very well with the theoretical PL relation from \cite{inayoshi2013} for the spherical accretion case of a bloated MYSO. In contrast, the longer period of 440--460 days (with an error of 70--80 days) is consistent with the prediction for the disc accretion case, which seems a more plausible scenario for I19520 because this source is known to be driving a collimated outflow.

\item
The source shows a negative slope in the optical CMD, which suggests the source is getting bluer while getting fainter. The observed trend disfavours the change of extinction along the line of sight, and hot and cold spots as a cause of variability. The intrinsic change in the source flux due to stellar pulsations, variable mass accretion rate, and variations in the inner disc structure could explain the observed colour trends. 

\end{itemize}

Overall, the nature of the variability in I19520, its observed period, and colour trend can be well explained by the theoretical pulsation model considering the bloated nature of this source and possible accretion through a disc. This is the first time that such a test has been carried out at optical and infrared wavelengths. While future variability studies at NIR wavelengths and spectroscopic time series analysis would reveal a more comprehensive picture for this crucial phase of high-mass star formation, it would be of decisive importance that theoretical models are further explored considering accretion through a disc and a broader range of effective temperatures. This would allow accurate comparisons of bloated MYSO candidates like I19520 with theoretical models, greatly improving our understanding of this unexplored phase of massive star formation.

\section*{Data availability}
The data underlying this article will be shared on reasonable request to the corresponding author.

\section{Acknowledgments}
\label{sec:ack}
We appreciate the anonymous referee's insightful and valuable feedback, which helped us improve the manuscript.
RP acknowledges Kei Tanaka, Vibhore Negi, Nikita Rawat, and Arpan Ghosh for an insightful and fruitful discussion. RP thanks UNAMDGAPA for a postdoctoral fellowship.
RP and AP acknowledge financial support from the UNAM-PAPIIT IG100223 grant. AP acknowledges financial support from the Sistema Nacional de Investigadores of CONAHCyT, M\'exico. 
AP, JS, JH, and CRZ acknowledge financial support from the CONAHCyT project number 86372 of the `Ciencia de Frontera 2019’ program, entitled `Citlalc\'oatl: A multiscale study at the new frontier of the formation and early evolution of stars and planetary systems’, M\'exico.
CSC acknowledges financial support through the I+D+i project PID2019-105203GB-C22 funded by the Spanish MCIN/AEI/10.13039/501100011033. 
RK acknowledges financial support via the Heisenberg Research Grant funded by the Deutsche Forschungsgemeinschaft (DFG, German Research Foundation) under grant no.~KU 2849/9, project no.~445783058. 
ASM acknowledges support from the RyC2021-032892-I grant funded by MCIN/AEI/10.13039/501100011033 and by the European Union `Next GenerationEU’/PRTR, as well as the program Unidad de Excelencia María de Maeztu CEX2020-001058-M, and support from the PID2020-117710GB-I00 (MCI-AEI-FEDER, UE). The Joan Or\'O Telescope (TJO) of the Montsec Observatory (OdM) is owned by the Catalan Government and opertayed by the Insititue of Space Studies of Catalonia (IEEC). 
RS's contribution to the research described here was carried out at the Jet Propulsion Laboratory, California Institute of Technology, under a contract with NASA, and funded in part by NASA via ADAP award/task order number 80NM0018F0610.
This work has made use of data from the European Space Agency (ESA) mission
{\it Gaia} (\url{https://www.cosmos.esa.int/gaia}), processed by the {\it Gaia}
Data Processing and Analysis Consortium (DPAC,
\url{https://www.cosmos.esa.int/web/gaia/dpac/consortium}). Funding for the DPAC
has been provided by national institutions, in particular, the institutions
participating in the {\it Gaia} Multilateral Agreement.


\bibliography{iras}
\bibliographystyle{mnras}



\appendix

\bsp    
\label{lastpage}

\end{document}